\documentclass[twocolumn,english,aps,prb,psfig,floats,showpacs]{revtex4-1}
\usepackage[latin9]{inputenc}
\usepackage{amsmath}
\usepackage{amssymb}
\usepackage{graphicx}
\usepackage{esint}
\usepackage[german]{babel} 

\makeatletter
\@ifundefined{textcolor}{}
{%
 \definecolor{BLACK}{gray}{0}
 \definecolor{WHITE}{gray}{1}
 \definecolor{RED}{rgb}{1,0,0}
 \definecolor{GREEN}{rgb}{0,1,0}
 \definecolor{BLUE}{rgb}{0,0,1}
 \definecolor{CYAN}{cmyk}{1,0,0,0}
 \definecolor{MAGENTA}{cmyk}{0,1,0,0}
 \definecolor{YELLOW}{cmyk}{0,0,1,0}
}

\usepackage{bm}\usepackage{latexsym}\usepackage{epsfig}

\makeatother

\usepackage{babel}
\begin{document}

\title{Field induced phase transitions in the helimagnet ${\bf Ba_{2}CuGe_{2}O_{7}}$}

\author{J. Chovan$^{1,*}$, M. Marder$^{2,\dag}$ and N. Papanicolaou$^{3,\ddag}$}

\affiliation{$^{1}$Department of Physics, Matej Bel University, Bansk{\'a} Bystrica,
Slovakia}

\email{Jaroslav.Chovan@umb.sk}

\selectlanguage{english}%

\affiliation{$^{2}$Center for Nonlinear Dynamics and Department of Physics, The
University of Texas at Austin, Austin, Texas 78712 }

\email{marder@mail.utexas.edu}

\selectlanguage{english}%

\affiliation{$^{3}$Department of Physics and Institute of Plasma Physics, University of Crete, Heraklion,
Greece}

\email{papanico@physics.uoc.gr}
\selectlanguage{english}%

\date{\today}
\begin{abstract}
\textbf{Abstract.} We present a theoretical study 
of the two-dimensional spiral antiferromagnet $\mathrm{Ba_{2}CuGe_{2}O_{7}}$
in the presence of an external magnetic field.  
We employ a suitable nonlinear $\sigma$ model to calculate
the $T=0$ phase diagram and the associated low-energy spin dynamics
for arbitrary canted magnetic fields, in general agreement with experiment.
In particular, when the field is applied parallel to the $c$ axis, 
a previously anticipated 
Dzyaloshinskii-type incommensurate-to-commensurate phase transition
is actually mediated by an intermediate phase, in agreement with our earlier
theoretical prediction confirmed by the recent observation of the
so--called double-$k$ structure.
The sudden $\pi/2$ rotations of the magnetic structures
observed in experiment are accounted for by a weakly
broken $U(1)$ symmetry of our model.
Finally, our analysis suggests a nonzero weak-ferromagnetic
component in the underlying Dzyaloshinskii-Moriya anisotropy, which
is important for quantitative agreement with experiment.
\end{abstract}


\maketitle
\vskip2pc

\section{Introduction}

\label{sec:intro} The presence of Dzyaloshinskii-Moriya (DM) anisotropy\cite{1,2}
in low-symmetry magnetic crystals typically leads to weak ferromagnetism,
as a result of slight spin canting in an otherwise antiferromagnetic
(AF) ground state. Another possibility is the occurrence of helimagnetism
whereby spins are arrayed in a helical or spiral structure whose period
(pitch) extends over several decades of unit cells.
These structures have intensively been studied lately, and the interest stems from a number of factors.
Some DM helimagnets display appealing magnetoelectric or multiferroic properties\cite{me1,me_vlx,me2}.
This enables the control of unusual magnetic states by electric fields and vice versa, and makes
these materials attractive for spintronics applications.
Another major factor is that, in addition to 1D spin spirals, 
the \emph{ground states} can form 
a vortex (skyrmion) lattice, as advocated by Bogdanov et al.\cite{vlt1,vlt2,vlt3} in a number of related models.
Recently, skyrmion-lattice ground states were observed experimentally 
in several magnetic systems\cite{me_vlx,vlx1,vlx2,vlx3} .  
Nontrivial types of localized \emph{nonlinear excitations} (domain walls) in DM helimagnets
have also been discussed in recent theoretical works\cite{dw1,dw2,dw3}.

Ba$_{2}$CuGe$_{2}$O$_{7}$
is an example of a helimagnet well
suited for experimental investigation thanks to a fortunate combination
of physical properties. It is an 
insulator whose magnetic properties
can be understood in terms of localized $s=\frac{1}{2}$ spins carried
by the Cu$^{2+}$ ions. The scale of energy set by an exchange constant
$J\sim1$ meV is very convenient for neutron scattering experiments.
Because of the low tetragonal symmetry (space group P$\bar{4}$2$_{1}$m)
the corresponding Heisenberg Hamiltonian involves an interesting combination
of antisymmetric (DM) as well as symmetric exchange anisotropies which
lead to a rich phase diagram. In particular, the strength of anisotropy
is such that magnetic phase transitions take place at critical fields
that are well within experimental reach.

A series of experiments in the late nineties\cite{3,4,5,6,7}
revealed the existence of a Dzyaloshinskii-type\cite{8} incommensurate-to-commensurate
(IC) phase transition when the strength of an external field applied
along the $c$ axis exceeds a critical value, $H_{c}\sim$2 T. For
$H<H_{c}$ the ground state is an incommensurate spiral whose period
$L=L(H)$ grows to infinity in the limit $H\rightarrow H_{c}$. For
$H>H_{c}$ the ground state was thought 
to become a commensurate 
antiferromagnet, a \emph{spin-flop} state. 
We note that the Dzyaloshinskii-type transition
is similar to the cholesteric-nematic phase transition induced by
an external magnetic field in chiral liquid crystals\cite{9,10,11}.
But the IC transition observed in Ba$_{2}$CuGe$_{2}$O$_{7}$ was the first
clean realization of the Dzyaloshinskii scenario in its original context,
and as such still remains very rare.

We carried out a detailed theoretical investigation\cite{12,13,14} inspired
by the earlier experimental work\cite{3,4,5,6,7} and predicted that the
IC phase transition does not occur immediately; instead,
between the incommensurate and commensurate phases occurs a separate
\emph{intermediate phase.} In short, there exist two critical fields,
$H_{c1}$ and $H_{c2}$ such that $H_{c1}<H_{c}<H_{c2}$ where $H_{c}\sim2$
T is the critical field for the presumed Dzyaloshinskii-type phase
transition. For $H<H_{c1}\sim1.7$ T the ground state is a flat spiral
(cycloid) that propagates along the $x$ axis while the staggered
magnetization rotates in the $xz$ plane. For $H>H_{c1}$ the cycloid
transforms into a nonflat spiral where all three components of the
staggered magnetization are different from zero.
Such a state may concisely be described as an antiferromagnetic conical
spiral that propagates along the $x$ axis while it nutates around
the $y$ axis. 
Above
$H_{c2}\sim2.9$ T the spiral becomes a conventional
commensurate antiferromagnet. 
This state is a commensurate antiferromagnetic spin-flop state
which is the ground state for all $H>H_{c2}$. Therefore the Dzyaloshinskii
field $H_{c}$ is not a true critical field, and the corresponding
IC phase transition is actually mediated by an additional phase in
the region $H_{c1}<H<H_{c2}$.

This prediction\cite{12} remained unexplored for almost a decade.
Additional experimental work on Ba$_{2}$CuGe$_{2}$O$_{7}$ has revealed its remarkable
magnetoelectric properties and demonstrated the electrical switching of magnetic propagation
vector and  the control of electric polarization by magnetic fields\cite{me1}.
However a new series of experiments has now confirmed the occurrence
of an intermediate phase in the form of an antiferromagnetic conical
spiral which has been called a \emph{double-$k$ structure} by the
experimental discoverers. 
This state occurs as predicted when an external
magnetic field is applied almost perfectly parallel to the $c$ axis\cite{16},
while further experiments have also explored the phase diagram
in the presence of an arbitrary canted magnetic field\cite{17}. 
It was not immediately evident to the experimentalists that
they had found what we predicted.
Our current task is to confirm that the recently observed double-$k$
structure is indeed the intermediate phase predicted in Ref.~24
and further to calculate the phase diagram in arbitrary canted magnetic
fields so as to complete the connection with the latest experiments.

In Sec.~\ref{sec:nl-sigma-model} we describe
the discrete spin Hamiltonian and its continuum approximation
in the form of a nonlinear $\sigma$ model.
In Sec.~\ref{sec:T=0-phase-diagram}, the 
ground state properties and the associated low-energy dynamics
will be calculated from the nonlinear $\sigma$ model
for a magnetic field of varying strength and direction. 
Hence, in Sec.~\ref{sec:Field-parallel-to}
the field is restricted to point along the $c$ (or $z$) axis and
its strength is varied through the IC transition.  
We recover then results of Ref.~24 
and further discuss the nature and stability of the intermediate phase.
An explicit calculation of the low-energy magnon spectrum throughout the intermediate phase,
and hence the opportunity for comparison with future experiments,  
is relegated to Appendix~A.
In Sec.~\ref{sec:Field-perpendicular-to} we study the case of a field applied
in a direction perpendicular to the $c$ axis. We thus recover an
experimentally observed bisection rule and further illuminate the
role of the out-of-plane DM anisotropy $d_{z}.$ The case of a magnetic
field applied in an arbitrary direction (canted magnetic field) is
analyzed in Sec.~\ref{sec:Canted-magnetic-fields} where we present
a theoretical prediction for the $T=0$ phase diagram, in fair agreement
with recent experiments. 
Local stability of the spin-flop phase in the
presence of arbitrary canted fields is shortly discussed in Appendix B.
Our main conclusions are summarized
in Sec.~\ref{sec:Conclusion}.

\section{Nonlinear $\sigma$ model}
\label{sec:nl-sigma-model}

In the method of calculation we closely follow the work of Ref.~24.
Ba$_{2}$CuGe$_{2}$O$_{7}$ is a layered compound
where the Cu atoms 
with spin $s=\frac{1}{2}$ 
form a perfect square lattice within each layer
with natural
axes $x$ and $y$ and lattice constant $l=5.986$ {\AA{}}. 
We note that the axes $x$, $y$ differ
from the conventional crystal axes $a$, $b$ by a $45^{\circ}$ azimuthal rotation  
The major spin interaction between in-plane neighbors is antiferromagnetic, 
while the interaction between out-of-plane neighbors is ferromagnetic and
weak. 
Therefore the interlayer coupling is ignored in the following
discussion which concentrates on the two-dimensional dynamics within
each layer.
  
The 2D spin Hamiltonian is of the general form
\begin{eqnarray}
 W & = & \sum_{<kl>} \left[ J_{kl}\left({\bf S}_{k}\cdot {\bf S}_{l}\right) +
{\bf D}_{kl}\cdot\left({\bf S}_{k}\times{\bf S}_{l}\right)\right] + \label{eq:1.1}\\
& & + \frac{1}{2}\sum_{<kl>}\sum_{i,j}G^{ij}_{kl}
         \left( S^{i}_{k}S^{j}_{l} + S^{j}_{k}S^{i}_{l} \right)
-\sum_{l}(g {\mu}_{B}{\bf H}\cdot{\bf S}_{l}),\nonumber
\end{eqnarray}
where ${\bf S}_{k}$ is the spin localized at site $k$, 
which satisfies the classical constraint ${\bf S}^2_{k}=s^2$.  
The first and the second terms in Eq.~(\ref{eq:1.1}) 
desribe the isotropic exchange interaction and antisymmetric
DM anisotropy over in-plane bonds denoted by $<kl>$.
The third term contains all symmetric exchange anisotropies, and the indices
$i$ and $j$ are summed over the three three values corresponding 
to the Cartesian components of the spin vectors along the axes $x$, $y$ and
$z$.
Single-ion anisotropy is not present in this spin $s=\frac{1}{2}$ system. 
Finally, the last term describes the usual Zeeman interaction with an
external field ${\bf H}$. 

The form of the interaction parameters is significantly restricted by the
crystal symmetry (space group P$\bar{4}$2$_{1}$m). 
It is safe to consider only nearest-neighbor (nn) in-plane bonds and
neglect interactions between next-nearest-neighbors\cite{12,13,14}.
Symmetry requires that the exchange constant $J = J_{kl}$  
is the same for
all nn in-plane bonds, whereas the constant vectors ${\bf D}_{kl}$ which account
for pure DM anisotropy are of the form 
\begin{eqnarray}
{\bf D}_{kl} & = & \left( 0, D_{\perp}, {\pm}D_z\right)
\hspace{0.2cm}\textrm{for bonds along} \hspace{0.2cm}x
\label{eq:1.2}\\
{\bf D}_{kl} & = & \left( D_{\perp}, 0, {\pm}D_z\right) 
\hspace{0.2cm}\textrm{for bonds along} \hspace{0.2cm}y,
\nonumber  
\end{eqnarray}
where $ D_{\perp}$, and ${\pm}D_z$ are two independent scalar constants.
It should be noted that the $z$-component of the DM vectors alternates
in sign on opposite bonds, a feature that could lead to weak ferromagnetism.
No such alternation occurs for the in-plane components of the DM vectors
which are responsible for the observed spiral magnetic order
or helimagnetism.

The symmetric exchange anisotropy will be restricted to the special KSEA limit\cite{ksea} 
throughout this paper. In this limit, the (traceless) symmetric tensor $G^{ij}_{kl}$
is expressed entirely in terms of the corresponding DM vector ${\bf D}_{kl}$ 
\begin{equation}
G^{ij}_{kl}=\frac{D^i_{kl}D^j_{kl}}{2J_{kl}}-\frac{|{\bf D}_{kl}|^2}{6J_{kl}}{\delta}_{ij}, \label{eq:1.3}
\end{equation}
where ${\delta}_{ij}$ is the Kronecker delta.
The KSEA limit has been shown to explain quantitatively a large set of experimental 
data\cite{5}, including some finer issues such as the lattice 
pinning of helical magnetic domains\cite{15}, and will be adopted here without further questioning. 
The Hamiltonian of Eq.~(\ref{eq:1.1}) is still consistent with the underlying space group P$\bar{4}$2$_{1}$m but
is not the most general Hamiltonian allowed by symmetry\cite{12,14}.
To our knowledge, Ba$_{2}$CuGe$_{2}$O$_{7}$  is the only known pure KSEA system. 
In this respect we mention that the layered antiferromagnet 
K$_{2}$V$_{3}$O$_{8}$ is not described by the KSEA anisotropy, as incorrectly stated in Ref.~31, 
because the observed easy-axis anisotropy is impossible to occur in the KSEA limit\cite{14}.

The discrete Hamiltonian of Eq.~(\ref{eq:1.1}) could be, 
in principle, 
analyzed by standard spin-wave techniques 
but such a task is technically complicated. 
The ground state 
and 
low-energy dynamics can be calculated 
from a simpler continuum field theory,  
which is a reasonable approximation because the period of
the observed magnetic spiral is
sufficiently long, about 37 lattice constants at zero field.
We omit technical details but stress the important steps
of the continuum approximation.

The major spin interaction is antiferromagnetic ($J=0.96$ meV) and
sets the energy scale of the system.
We therefore divide a complete magnetic 
lattice into two sublattices A, B and then 
rewrite the Landau-Lifshitz equation as a system of 
two coupled equations for the sublattice spins 
${\bf A}$, ${\bf B}$. 
However, a more transparent formulation is obtained in terms of new variables, 
the magnetization ${\bf m}=({\bf A}+{\bf B})/2s$ 
and the staggered magnetization ${\bf n}=({\bf A}-{\bf B})/2s$, which satisfy
the classical constraints ${\bf m}\cdot{\bf n} = 0$ and ${\bf m}^2 + {\bf n}^2 = 1$.
The basis for the derivation of an effective field theory is the fact
that all anisotropies and the applied field
$D_{\perp}$, $D_{z}$, $g {\mu}_{B} H/s$ are significantly smaller than
the exchange constant $J$. Consequently, 
$|{\bf m}|$ is also much smaller than $|{\bf n}|$, and both
${\bf m}$, ${\bf n}$ vary appreciably only over distances of
many lattice spacings.

  To ascertain the relative significance of the various terms 
that arise during a consistent low-energy reduction, one may
employ a dimensionless scale $\varepsilon$ 
defined from, say, ${\varepsilon} = D_{\perp}/J$.
We further introduce rescaled (dimensionless) anisotropy 
$d_z=\sqrt{2}D_z/{\varepsilon}J$ 
and magnetic field 
${\bf h} = g {\mu}_{B}{\bf H}/(2\sqrt{2}s{\varepsilon}J)$.
Note that the unit of field ($h=1$) corresponds to  $2\sqrt{2}s{\varepsilon}J/g {\mu}_{B}=1.68$ T,
where we use the values $s=1/2$, $g=2.474$, $J=0.96$ meV and $\varepsilon=0.1774$ thought
to be appropriate for the description of Ba$_{2}$CuGe$_{2}$O$_{7}$.
Similarly, we introduce rationalized spatial coordinates $x$, $y$ and time $t$,
and complete our choice with the statement that 
frequency is measured in units of $\hbar\omega=2\sqrt{2}s{\varepsilon}J=0.24$ meV,
and distance in units of $l/{\varepsilon}=33.75$ {\AA{}},
where $l$ is the lattice constant of the square lattice formed by the Cu atoms.
 
The continuum approximation is then obtained by a systematic 
formal expansion of Landau-Lifshitz equation in powers of $\varepsilon$,
where both ${\bf m}$, ${\bf n}$ 
are considered as continuous functions
of the (dimensionless) in-plane spatial coordinates $x$, $y$.
The magnetization ${\bf m}$ is treated 
as a quantity of order $\varepsilon$, whereas the staggered magnetization ${\bf n}$
and the rescaled variables are assumed to be of order of unity.
Then, to leading order, the classical constraints reduce to 
${\bf m}\cdot{\bf n} = 0$ and ${\bf n}^2 = 1$.
Finally, the $T=0$ low-energy dynamics is expressed entirely 
in terms of the staggered magnetization ${\bf n}$, 
and is calculated from a  
nonlinear $\sigma$ model with Lagrangian density ${\cal L}$:

\begin{eqnarray}
{\mathcal{L}} & = & {\mathcal{L}}_{0}-V;\label{eq:1.4}\\
{\mathcal{L}}_{0} & = & \frac{1}{2}\partial_{0}{\bf n}\cdot\partial_{0}{\bf n}+{\bf h}\cdot{\bf n}\times\partial_{0}{\bf n};\nonumber \\
V & = & \frac{1}{2}\left(\partial_{1}{\bf n}-{\bf e}_{2}\times{\bf n}\right)^{2}
+\frac{1}{2}\left(\partial_{2}{\bf n}-{\bf e}_{1}\times{\bf n}\right)^{2}\nonumber \\
 &  & +\frac{1}{2}({\bf n}\cdot{\bf h})^{2}+d_{z}({\bf h}\times{\bf e}_{3})\cdot{\bf n}.\nonumber 
\end{eqnarray}
Here ${\bf e}_{1}$, ${\bf e}_{2}$, and ${\bf e}_{3}$ are unit vectors
along the $x,$ $y,$ and $z$ axes, whereas derivatives
are described by $\partial_{1}=\partial/\partial x,$ $\partial_{2}=\partial/\partial y,$
and $\partial_{0}=\partial/\partial t.$
The applied magnetic field
${\bf h}=h_{1}{\bf e}_{1}+h_{2}{\bf e}_{2}+h_{3}{\bf e}_{3}$ may
point in any arbitrary direction. 
We emphasize that the staggered magnetization 
${\bf n}=n_{1}{\bf e}_{1}+n_{2}{\bf e}_{2}+n_{3}{\bf e}_{3}$
is a \emph{unit vector field} (${\bf n}^2=1$)
that depends
upon the in-plane spatial coordinates $x$ and $y$ as well as the
time variable $t$: ${\bf n}={\bf n}(x,y,t)$.
The in-plane component of the DM anisotropy $D_{\perp}$
has been completely suppressed in Eq.~(\ref{eq:1.4})
through the definions of rationalized units.

It should be noted that the special KSEA limit adopted here
is equivalent to setting $\kappa=0$ in the Lagrangian of Ref.~24. 
Otherwise Eq.~(\ref{eq:1.4}) gives the most general Lagrangian 
compatible with symmetry, expressed in fully rationalized units.  
Rationalized units greatly simplify the analysis of  Eq.~(\ref{eq:1.4})
and will be employed throughout our theoretical 
development in the remainder of the paper.
However, the critical or otherwise significant values of the field
will occasionaly be quoted also in physical units,
in order to facilitate the orientation of the reader and comparison
with experiment. 
Essentially for the same reason we use physical units  
in all figures directly relevant to experiment \cite{16,17}.

\section{T=0 phase diagram}
\label{sec:T=0-phase-diagram}

\begin{figure*}
\centering \resizebox{0.95\textwidth}{!}{\includegraphics{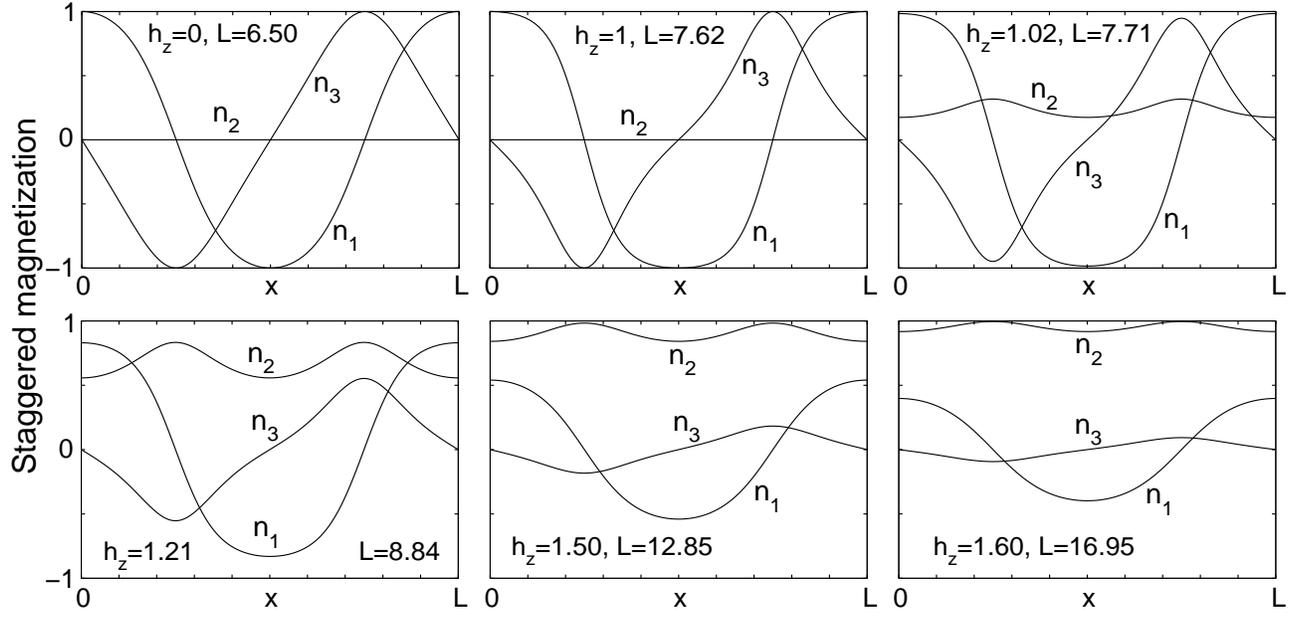}
} 
\caption{Solutions of Eqs.~(\ref{eq:2.5}) minimizing the average energy density of Eq.(\ref{eq:2.13}) for a number
of illustrative magnetic fields $h=h_z$ pointing along the $c$ axis. Note that the period $L$ varies
with the field. Above $h_{c2}=\sqrt{3}$ the solution becomes a commensurate antiferromagnet with $n_2 = 1$,
$n_1 = n_3 = 0$. }

\label{fig:1} 
\end{figure*}

\begin{figure}
\resizebox{0.47\textwidth}{!}{\includegraphics{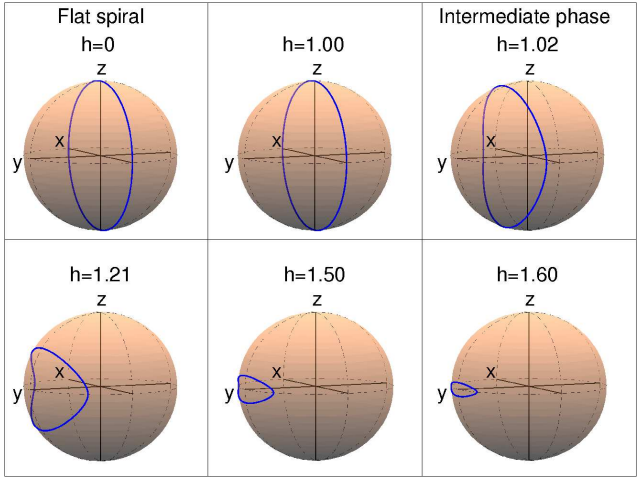} } 
\caption{ (Color online) The same solutions as in Fig.~1, but viewed
from a different perspective. 
Blue lines on the sphere surface trace out directions for the staggered
magnetization, placing the base of each ${\bf n}$ at the center of
the sphere and then moving from one unit cell to the next along $x$,
the direction of spiral propagation. For $0<h<h_{c1}$ the spins describe
a cycle in the $xz$ plane, while for $1<h<h_{c2}=\sqrt{3}$ they
also have a nonzero oscillating $y$ component.}

\label{fig:2} 
\end{figure}

\subsection{Field parallel to $c$}
\label{sec:Field-parallel-to} We begin by specializing to the case
where the magnetic field is applied strictly along the $c$ axis:
${\bf h}=h{\bf e}_{3}$. Then the potential $V$ of Eq.~(\ref{eq:1.4})
reduces to 
\begin{eqnarray}
V & = & \frac{1}{2}\left[\left(\partial_{1}{\bf n}\right)^{2}+\left(\partial_{2}{\bf n}\right)^{2}+\left(1+h^{2}\right)n_{3}^{2}+1\right]\nonumber \\
 &  & -\left[\left(\partial_{1}n_{1}-\partial_{2}n_{2}\right)n_{3}-\left(n_{1}\partial_{1}-n_{2}\partial_{2}\right)n_{3}\right]\,
\end{eqnarray}
and is symmetric under the $U(1)$ transformation 
\begin{equation}
x+iy\rightarrow(x+iy)\ e^{i{\psi}},\,\,\,\, n_{1}+in_{2}\rightarrow(n_{1}+in_{2})\ e^{-i{\psi}},\label{eq:2.1}
\end{equation}
which is somewhat unusual in that an azimuthal rotation of spatial
coordinates $x$ and $y$ by an angle $\psi$ is followed by a corresponding
rotation of the staggered magnetization by an angle $-\psi$.

The ground state is obtained by finding energy-minimizing solutions
${\bf n}$ of the static energy functional 
\begin{equation}
W=\int dx\, dy\: V.\label{eq:2.2}
\end{equation}
In order to enforce the constraint that the staggered magnetization
be of unit length we adopt a parameterization

\begin{equation}
{\bf n}=\sin\Phi\sin\Theta\,{\bf e}_{1}+\cos\Theta\,{\bf e}_{2}+\cos\Phi\sin\Theta\,{\bf e}_{3},\label{eq:2.3}
\end{equation}
which differs from more standard parameterizations by a circular permutation,
but turns out to yield slightly more compact expressions later on.

Our first task is to find solutions $\Theta=\Theta(x,y),\ \Phi=\Phi(x,y)$
that minimize $W$. 
The minimum of energy is sought after in the form of the one-dimensional (1D)
Ansatz 
\begin{equation}
\Theta(x,y)=\theta(x);\quad\Phi(x,y)=\phi(x).\label{eq:2.4}
\end{equation}
which assumes that the staggered magnetization depends only
on the spatial coordinate $x$.
In view of the $U(1)$ symmetry in Eq.~(\ref{eq:2.1}) any solution
we find of this type automatically produces a family of additional
solutions of the same energy rotated by angle $\psi.$ Varying $W$
then yields 
\begin{eqnarray}
\partial_{1}^{2}\phi & = & -\frac{\left(2\,\partial_{1}\phi-2\right)\,\cos\theta\,\partial_{1}\theta+\gamma^{2}\,\cos\phi\,\sin\phi\,\sin\theta}{\sin\theta}\nonumber \\
\nonumber \\
\partial_{1}^{2}\theta & = & \left(\left(\partial_{1}\phi\right)^{2}-2\,\partial_{1}\phi+\gamma^{2}\,\cos^{2}\phi\right)\,\cos\theta\,\sin\theta.\label{eq:2.5}
\end{eqnarray}
with $\gamma^{2}=1+h^{2}.$ Here subscript $1$ indicates a derivative
with respect to $x$. All derivatives with respect to 
$y$ vanish because we are working in a space of one-dimensional
solutions.

To illustrate the solutions, we first consider the special case of
a \emph{flat spiral} (cycloid) with $\theta=\pi/2$. Then the second
of Eqs.~(\ref{eq:2.5}) is automatically satisfied and the first becomes
\begin{equation}
\partial_{1}^{2}\phi+\gamma^{2}\cos\phi\sin\phi=0,\label{eq:2.6}
\end{equation}
while the staggered magnetization becomes 
\begin{equation}
{\bf n}=(\sin\phi,0,\cos\phi),\label{eq:2.7}
\end{equation}
a cycloid that propagates along the $x$ axis while rotating in the
$xz$ plane (upper left panel of Fig.~\ref{fig:1} and Fig.~\ref{fig:2}). The solution
for $\phi$ obeys 
\begin{equation}
\partial_{1}\phi=\sqrt{\delta^{2}+\gamma^{2}\cos^{2}\phi},\quad x=\int_{0}^{\phi}\frac{d\varphi}{\sqrt{\delta^{2}+\gamma^{2}\cos^{2}\varphi}}.\label{eq:2.8}
\end{equation}
The result can be expressed in terms of elliptic functions but there
is no particular advantage to doing so. $\delta^{2}$ is a positive
constant that will be determined below. The cycloid has a period (pitch)
of 
\begin{equation}
L=\int_{0}^{2\pi}\frac{d\phi}{\sqrt{\delta^{2}+\gamma^{2}\cos^{2}\phi}}\label{eq:2.9}
\end{equation}
and the free parameter $\delta$ is determined by the requirement
that the average energy density $w=W/L$ achieve a minimum: 
\begin{equation}
\frac{1}{2\pi}\int_{0}^{2\pi}d\phi\sqrt{\delta^{2}+\gamma^{2}\cos^{2}\phi}=1\Rightarrow w=\frac{1}{2}(1-\delta^{2}).\label{eq:2.10}
\end{equation}
As $\gamma$ (or $h$) increases $\delta$ becomes zero at a critical
field: 
\begin{equation}
\gamma=\gamma_{c}=\pi/2\Rightarrow h=h_{c}=\sqrt{\frac{\pi^{2}}{4}-1}\approx1.21.\label{eq:2.11}
\end{equation}
In physical units, $H_{c}=2.04$ T. This is the Dzyaloshinskii critical
field and the corresponding Dzyaloshinskii scenario may be described
as follows: for $h<h_{c}$ the solution is a flat spiral that propagates
along the $x$ axis and rotates in the $xz$ plane. As $h$ approaches
$h_{c}$ the spiral is highly distorted and becomes a kink-like structure
with diverging period. For $h>h_{c}$ the ground state becomes the
uniform spin-flop state 
\[
{\bf n}=(1,0,0)\quad\text{modulo}\ U(1).
\]
We realized that this scenario was incomplete when we computed the
magnon spectrum \cite{12} of the flat spiral and found negative eigenvalues
starting at 
\begin{equation}
h_{c1}=1.01,\quad H_{c1}=1.7\text{ T.}\label{eq:2.12}
\end{equation}
The fact that the value of $h_{c1}$ in rationalized units  
is practically equal to 1 is remarkable, yet fortuitous
and bears no special significance otherwise.
Above $h_{c1}$ the flat spiral is unstable.
We thus return to energy
minimization and revoke the assumption $\theta=\pi/2,$ although continuing
to assume a one-dimensional structure of the form $\phi=\phi(x)$
and $\theta=\theta(x)$.

Efforts to find explicit analytical solutions of Eqs.~(\ref{eq:2.5})
have not been fruitful so we resort to numerics. We minimize the energy
density 
\begin{equation}
w=\frac{1}{L}\int_{0}^{L}dx\, V(\theta,\phi)\label{eq:2.13}
\end{equation}
over a periodic chain of length $L$ and vary $L$ to achieve a minimum
for any given value of $h$.

For $h<h_{c1}=1.01$ we recover the previous results for the flat
spiral. But for $h>h_{c1}$ a nonflat spiral arises with nontrivial
$\phi(x)$ as well as $\theta(x)$. 
We call this the \emph{intermediate state}.
Examples appear in Figs.~\ref{fig:1} and ~\ref{fig:2} for
a variety of field values. Entering the intermediate phase for $h>h_{c1}$,
$n_{2}$ acquires nonzero values and one can describe the state as
an antiferromagnetic conical spiral that propagates along $x$ but
nutates around $y$. 
This is precisely the structure deduced from recent
scattering experiments \cite{16,17} and called
a \emph{double-k structure} because of two-fold peak
characteristically observed during experimental scans through $k$ space. 
As $h$ increases, the component $n_{2}$ becomes
larger and larger until at $h_{c2}=\sqrt{3}$ (or $H_{c2}=2.9$ T)
the solution becomes 
a commensurate antiferromagnet or spin-flop state 
with ${\bf n}=(0,1,0).$ This
upper critical point was determined in Ref.~24 from a stability
analysis. The existence of the intermediate
state does not depend upon the presence of a nonzero transverse magnetic
field.
\\
\\
We now comment on the two issues
concerning the nature of the ground state for $h{\parallel}c$. 
Our first comment concerns the possible existence
of more general structures with average energy density lower
than those calculated above through the 1D Ansatz (\ref{eq:2.4}). 
In particular, ground states in the form of a vortex (skyrmion) lattice
in 2D Dzyaloshinskii-Moriya helimagnets have been speculated theoretically\cite{vlt1,vlt2}
(also in connection with Ba$_2$Cu$_2$O$_7$\cite{vlt3}),
and later discovered experimentally in several such systems\cite{me_vlx,vlx1,vlx2,vlx3}.
Hence, we carried out extensive two-dimensional simulations, but our numerical
investigation yielded negative results for a potential ground state in the form of, say,
a vortex lattice. 
Instead, in all our 2D numerical experiments, 
we found that the optimal configuration for $h_{c1}< h < h_{c2}$ 
is actually the same 1D nonflat spiral, 
which were obtained earlier in this Section
by a numerical minimization applied directly to a 1D restriction of the energy functional (\ref{eq:2.13}).

Second, the nonflat spiral, calculated numerically through the relaxation
algorithm, exists as a stationary point of the energy
functional in the region  $h_{c1} < h < h_{c2}$.
It is thus desirable to examine also its stability, and check 
whether or not there exist yet another critical field
within the intermediate region, beyond
which the nonflat spiral may cease to be locally stable.
We have therefore calculated the magnon spectrum of the intermediate phase
(Appendix A) and verified that all eigenvalues are always positive.
Consequently, the nonflat spiral is locally stable within the entire intermediate region.
This computation does not prove it is the ground state, but in combination
with extensive numerical explorations of two-dimensional states that
found no solutions of lower energy, it is a strong indication.
Note that the magnon spectrum in the intermediate phase has not been experimentally investigated yet. 
Hence, our current theoretical predictions provide the opportunity for comparison with future 
inelastic neutron scattering studies.

\subsection{Field perpendicular to $c$}

\label{sec:Field-perpendicular-to} We next consider a field applied
in a direction strictly perpendicular to the $c$ axis, a case that
had attracted experimental interest already in Ref.~18. For the
moment, we assume that the field is applied along the $y$ axis, ${\bf h}=(0,h_{\perp},0)$,
hence the potential $V$ of Eq.~(\ref{eq:1.4}) reduces to 
\begin{eqnarray}
V & = & \frac{1}{2}\left[\left(\partial_{1}{\bf n}\right)^{2}+\left(\partial_{2}{\bf n}\right)^{2}+n_{3}^{2}+h_{\perp}^{2}n_{2}^{2}+1\right]+h_{\perp}d_{z}n_{1}\nonumber \\
 &  & -\left[\left(\partial_{1}n_{1}-\partial_{2}n_{2}\right)n_{3}-\left(n_{1}\partial_{1}-n_{2}\partial_{2}\right)n_{3}\right]\,,\label{eq:3.1}
\end{eqnarray}
where the applied field enters in two distinct ways; namely through
the appearance of an effective easy-plane anisotropy $\frac{1}{2}h_{\perp}^{2}n_{2}^{2}$
and a Zeeman-like anisotropy $h_{\perp}d_{z}n_{1}$. The latter also
contains the strength $d_{z}$ of the out-of-plane oscillating component
$\left(\pm D_{z}\right)$ of the DM vectors which was neglected in
the analysis of Ref.~18.

To find minima of the energy functional we first note that the positive
term $\frac{1}{2}h_{\perp}^{2}n_{2}^{2}$ again favors a flat-spiral
configuration with $n_{2}=0$ which propagates along the $x$ axis.
Using the angular parametrization (\ref{eq:2.3}) we write 
\begin{equation}
\Phi=\phi\left(x\right)\,,\quad\Theta=\frac{\pi}{2}\,;\quad{\bf n}=\left(\sin\phi,\,0,\,\cos\phi\right)\,,\label{eq:3.2}
\end{equation}
which is inserted in Eq.~(\ref{eq:3.1}) to yield 
\begin{equation}
V=\frac{1}{2}\left[\left(\partial_{1}\phi-1\right)^{2}+\cos^{2}\phi\right]+\bar{h}\sin\phi\,,\label{eq:3.3}
\end{equation}
where the only free parameter 
\begin{equation}
\bar{h}=h_{\perp}d_{z}\label{eq:3.3a}
\end{equation}
is a combination of the applied field $h_{\perp}$ and the effective
out-of-plane DM anisotropy $d_{z}$.

Otherwise, the calculation is similar to that of the flat spiral in
Sec. \ref{sec:Field-parallel-to}. Stationary points of the energy functional $W=\int Vdx$ now
satisfy the ordinary differential equation 
\begin{equation}
\partial_{1}^{2}\phi+\cos\phi\sin\phi-\bar{h}\cos\phi=0\label{eq:3.4}
\end{equation}
whose first integral is given by 
\begin{equation}
\left(\partial_{1}\phi\right)^{2}-{\cos}^{2}{\phi}-2\bar{h}{\sin}{\phi}=C=2\bar{h}+{\delta}^{2}\,.\label{eq:3.5}
\end{equation}
Our choice of the integration constant $C$ indicates that minimum
energy is achieved with a positive new constant denoted by ${\delta}^{2}$.
The actual configuration $\Phi=\phi\left(x\right)$ is then given
by the implicit equation 
\begin{equation}
x=\int_{0}^{\phi}\frac{d\varphi}{\sqrt{\delta^{2}+{\cos}^{2}{\varphi}+2\bar{h}(1+{\sin}{\varphi})}}\,,\label{eq:3.6}
\end{equation}
and the corresponding spiral period $L$ is given by 
\begin{equation}
L=\int_{0}^{2\pi}\frac{d\phi}{\sqrt{\delta^{2}+{\cos}^{2}\phi+2\bar{h}(1+{\sin}{\phi})}}\,.\label{eq:3.7}
\end{equation}
Finally, the free parameter $\delta^{2}$ is calculated by minimizing
the average energy density $w=\frac{1}{L}\int_{0}^{L}V(x)dx$ which
yields 
\begin{equation}
{\frac{1}{2\pi}}\int_{0}^{2\pi}d{\phi}{\sqrt{\delta^{2}+{\cos}^{2}\phi+2\bar{h}(1+{\sin}{\phi})}}=1\,,\label{eq:3.8}
\end{equation}
an algebraic equation that may be used to determine $\delta^{2}$
for each value of $\bar{h}$. The corresponding minimum energy is
then given by 
\begin{equation}
w={\frac{1}{2}}(1-{\delta}^{2}-2\bar{h}).\label{eq:3.9}
\end{equation}

In the absence of the out-of-plane DM anisotropy $(d_{z}=0)$ the
configuration just calculated reduces to the zero-field flat spiral
of Sec. \ref{sec:Field-parallel-to} for any value of the applied transverse field because $\bar{h}=h_{\perp}d_{z}=0$
for all $h_{\perp}$. In particular, no phase transition of the Dzyaloshinskii
type would be expected to occur for a field applied in a direction
strictly perpendicular to the $c$ axis, as presumed in the analysis
of early experiments \cite{6}.

\begin{figure}
\resizebox{0.45\textwidth}{!}{\includegraphics{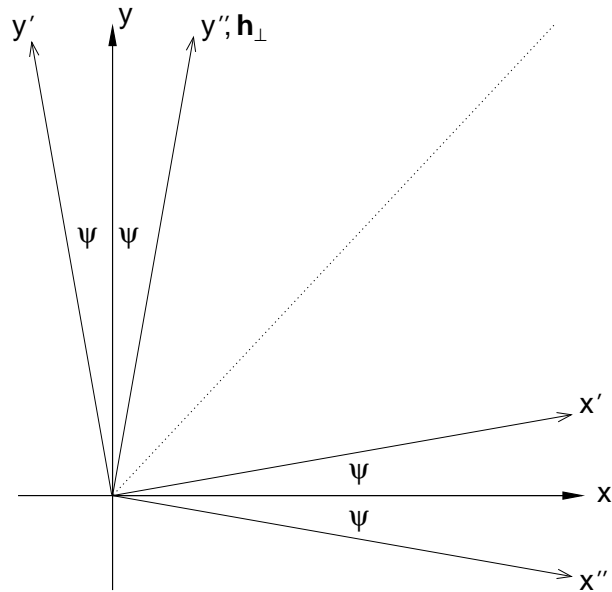} } 
\caption{Illustration of the bisection rule in a transverse field applied along
the $y^{\prime\prime}$ axis. The spiral propagation vector points
along the $x^{\prime}$ axis, while the staggered magnetization rotates
in the $x^{\prime\prime}z$ plane. }

\label{fig:3} 
\end{figure}

However, the situation changes significantly for $d_{z}\neq0$. Then
the effective field $\bar{h}=h_{\perp}d_{z}$ is different from zero
except when $h_{\perp}=0$. With increasing $h_{\perp}$, and thus
increasing $\bar{h}$, the parameter $\delta^{2}$ decreases and eventually
vanishes when $\bar{h}$ reaches a critical value $\bar{h}=\bar{h}_{c}$
computed from Eq.~(\ref{eq:3.7}) applied for $\delta^{2}=0$. A simple
numerical calculation yields $\bar{h}_{c}=h_{\perp}^{c}d_{z}=0.3161$,
or 
\begin{equation}
h_{\perp}^{c}=\frac{0.3161}{d_{z}}\,.\label{eq:3.10}
\end{equation}
In the limit $h_{\perp}\rightarrow h_{\perp}^{c}$, $\delta^{2}$
vanishes and the average energy density of Eq.~(\ref{eq:3.9}) reduces
to $w={\frac{1}{2}}(1-2\bar{h}_{c})$ which coincides with the energy
of the uniform spin-flop state ${\bf n}=(-1,0,0)$. Thus we again
encounter a Dzyaloshinskii-type phase transition at a critical field
that now depends on $d_{z}$.

As mentioned already, no such transition was detected in the early
experiments \cite{6} which were conducted with transverse magnetic
fields of limited strength $H_{\perp}\lesssim2$ T or $h_{\perp}\lesssim2/1.68\approx1.2$.
However, recent experiments \cite{17} reveal a critical field $H_{\perp}^{c}=9$
T or $h_{\perp}^{c}=9/1.68=5.36$ and, using Eq.~(\ref{eq:3.10}),
\begin{equation}
d_{z}=0.06.\label{eq:3.11}
\end{equation}
As far as we know, this is the first estimate of the strength of the
out-of-plane DM anisotropy and will be used in all numerical calculations
presented in the continuation of this paper. Incidentally, using the
definition of the rationalized anisotropy $d_{z}={\sqrt{2}}D_{z}/\varepsilon J$
from Ref.~24, we find $D_{z}/J=0.0076$, to be compared with
$\varepsilon=D_{\perp}/J=0.18$.

The preceding calculation was completed in Ref.~25 with a detailed
calculation of the corresponding magnon spectrum which could prove
useful for the analysis of future inelastic neutron scattering experiments
in the presence of a strong transverse magnetic field $H_{\perp}$.
The same calculation reveals no sign of further critical instabilities
as long as $d_{z}<0.5$. In particular, an intermediate phase of the
type encountered in Sec. \ref{sec:Field-parallel-to} is not present in the case of strictly
transverse magnetic fields and $d_{z}<0.5$. In other words, the predicted
phase transition is of pure Dzyaloshinskii type \cite{8}.

This section is completed with a brief discussion of the case of a
transverse magnetic field 
\begin{equation}
{\bf h}_{\perp}=h_{\perp}\left(\sin\psi,\cos\psi,0\right)\label{eq:3.12}
\end{equation}
which points in an arbitrary direction within the basal plane obtained
by a clockwise rotation of the $y$ axis with angle $\psi$ (see Fig.~3).
In fact, the ground-state configuration for this more general case
($\psi\neq0$) can be surmised from the special $\psi=0$ solution
calculated earlier in this section by simple algebraic transformations,
thanks to the underlying $U(1)$ symmetry of Eq.~(\ref{eq:2.1}) broken
by the applied transverse field. Indeed, let $n_{1}=n_{1}(x),n_{2}=0,n_{3}=n_{3}(x)$
be the $\psi=0$ solution. Then the solution for $\psi\neq0$ is given
by 
\begin{equation}
n_{1}^{\prime}=\cos\psi\, n_{1}(\xi)\,,\,\,\,\, n_{2}^{\prime}=-\sin\psi\, n_{1}(\xi)\,,\,\,\,\, n_{3}^{\prime}=n_{3}(\xi)\,,\label{eq:3.13}
\end{equation}
where $\xi=x\cos\psi+y\sin\psi$. Thus the new spiral propagates along
the $x^{\prime}$ axis obtained by a counter-clockwise rotation of
the $x$ axis with angle $\psi$ (see Fig.~3) while the staggered
magnetization rotates in the plane $x^{\prime\prime}z$ which is perpendicular
to the field direction (axis $y^{\prime\prime}$). In other words,
a flat spiral (cycloid) that initially propagates along the $x$
axis and rotates in the $xz$ plane ($\psi=0$) is reoriented to propagate
along the $x^{\prime}$ axis ($\psi\neq0$) so that the normal to
the spin plane (axis $y^{\prime\prime}$) points along the applied
magnetic field. The angle formed by the direction of spiral propagation
(axis $x^{\prime}$) and the normal to the spin plane (axis $y^{\prime\prime}$)
is bisected by the conventional crystal axis $b=(0,1,0)$ denoted
by a dotted line in Fig.~3 for any $\psi$. When the field is applied
along $b$, $\psi=\frac{\pi}{4}$ and the normal to the spin-rotation
plane is parallel to the propagation vector (screw-type spiral).

\begin{figure*} 
\resizebox{0.98\textwidth}{!}{\includegraphics{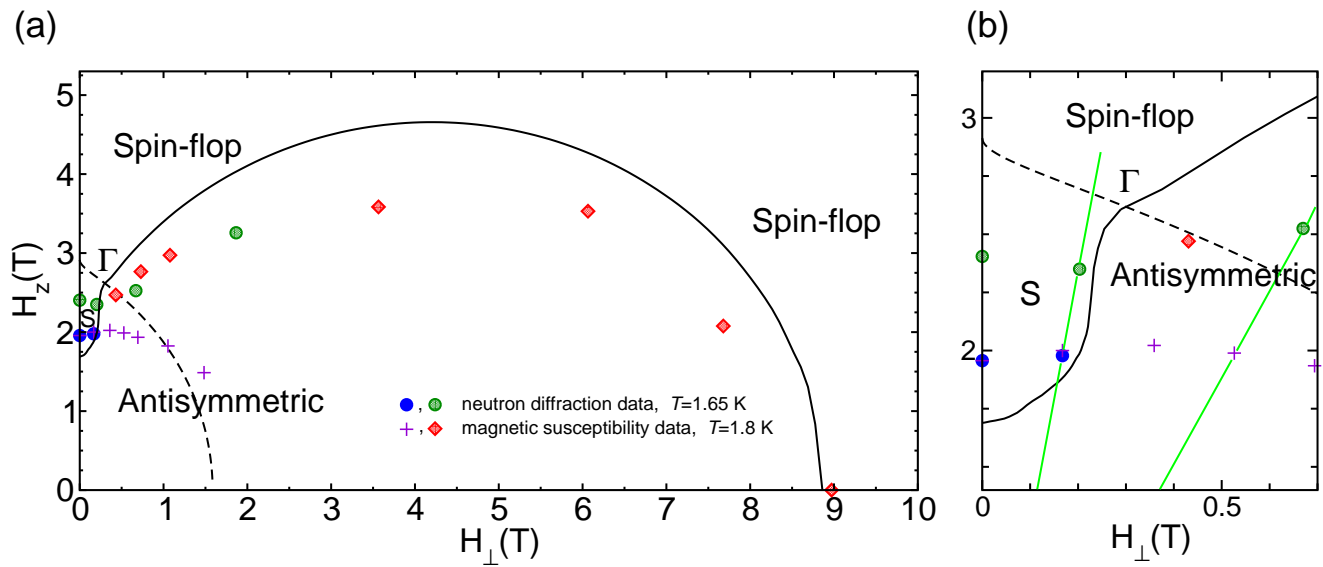} } 
\caption{(Color online) $T=0$ theoretical prediction for the phase diagram. 
We adopt conventions used in publication of experiments \protect\cite{17}.
The Antisymmetric and Symmetric phases reported here are illustrated
in Figures 5-6 and 7-8, respectively.
(a) The Antisymmetric
phase is realized below the solid line. The Symmetric phase, denoted
as S in the figure, exists in the area between the solid line and
the dashed line. The dashed line depicts the limit of local stability
of the Spin-flop phase. The Spin-flop phase is locally stable above
the dashed line, but is actually realized only in the area above both
the dashed line and the solid line. 
(b) A portion of the phase diagram near 
the tricritical point $\Gamma$ where the three phases 
(Symmetric, Antisymmetric, Spin-flop) merge.
Experimental data were extracted from Fig.~11(a) of Ref.29.
The straight solid (green) lines correspond to experimental
scans along magnetic field that will be discussed in the paragraph
on neutron scattering.
}

\label{fig:4} 
\end{figure*}

The ``bisection rule'' just described theoretically was experimentally
discovered already in Ref.~18. Actually, agreement with the ideal
bisection rule requires that $H_{\perp}\gtrsim0.5$ T in order to
overcome a certain energy barrier due to discreteness effects which
lead to an additional tetragonal anisotropy that breaks the underlying
$U(1)$ symmetry even in the absence of a transverse field \cite{6,15}.
The same anisotropy explains the experimental fact that the spiral
propagates along the $x=(1,1,0)$ or $x=(1,\bar{1},0)$ directions,
in the absence of a transverse field, while a sufficiently strong
field $H_{\perp}\gtrsim0.5$ T is required to reorient the spiral
according to the bisection rule.

We have thus completed the discussion of the phase diagram in the
presence of a field strictly parallel to the $c$ axis (Sec.~\ref{sec:Field-parallel-to}) or
a field strictly perpendicular to $c$ (Sec.~\ref{sec:Field-perpendicular-to}). The general case
of a canted magnetic field is discussed in the following Sec.~\ref{sec:Canted-magnetic-fields}.

\subsection{Canted magnetic fields}

\label{sec:Canted-magnetic-fields} We now turn our attention to the
most general case of the applied field ${\bf h}$, whose transverse
component $h_{\perp}$ and the component $h_{z}$ along the $c$ axis
are both nonzero. 
This is necessary to consider because experimentalists 
have reported ground state information on the system while
scanning through all field components.
For a while we assume that the magnetic field ${\bf h}$
is given by 
\begin{equation}
{\bf h}=h_{\perp}{\bf e}_{2}+h_{z}{\bf e}_{3}\,.\label{eq:4.1}
\end{equation}
The explicit form of the potential of Eq.~(\ref{eq:1.1}) becomes 
\begin{eqnarray}
V & = & \frac{1}{2}\left[\left({\partial}_{1}{\bf n}\right)^{2}+\left({\partial}_{2}{\bf n}\right)^{2}+1\right]-\label{eq:4.2}\\
 &  & -\left[\left(\partial_{1}n_{1}-\partial_{2}n_{2}\right)n_{3}-\left(n_{1}\partial_{1}-n_{2}\partial_{2}\right)n_{3}\right]\,\nonumber \\
 &  & +\frac{1}{2}\gamma^{2}n_{3}^{2}+\frac{1}{2}h_{\perp}^{2}n_{2}^{2}+h_{\perp}h_{z}n_{2}n_{3}+h_{\perp}d_{z}n_{1}\,,\nonumber 
\end{eqnarray}
where the parameter $\gamma^{2}$ depends upon $h_{z}$, 
\begin{equation}
\gamma^{2}=1+h_{z}^{2}.\label{eq:4.3}
\end{equation}

When $h_{\perp}\neq0$, a brief inspection of the potential of Eq.~(\ref{eq:4.2})
reveals that the Zeeman energy $\frac{1}{2}\left({\bf n}\cdot{\bf h}\right)^{2}$
now contains also the \emph{off-diagonal} anisotropy $h_{\perp}h_{z}n_{2}n_{3}$,
which was absent when either $h_{\perp}=0$ or $h_{z}=0$. The presence
of the latter anisotropy precludes analytical treatment. We therefore
obtain the corresponding solutions by a direct minimization of the
energy functional, in a manner analogous to the calculation presented
in Sec.~\ref{sec:Field-parallel-to}.
For the sake of clarity, we also recall 
the value of the out-of-plane DM anisotropy $d_z=0.06$ estimated in 
Sec.~\ref{sec:Field-perpendicular-to}, which is used
in all subsequent numerical calculations.
We state our $T=0$ results in the phase diagram in Fig.~\ref{fig:4}.
For comparison, we include experimental critical lines 
determined from neutron diffraction and magnetic suceptibility
measurements taken, however, at relatively high temperature $T=1.65$ K and
$1.8$ K.

\begin{figure}
\resizebox{0.48\textwidth}{!}{\includegraphics{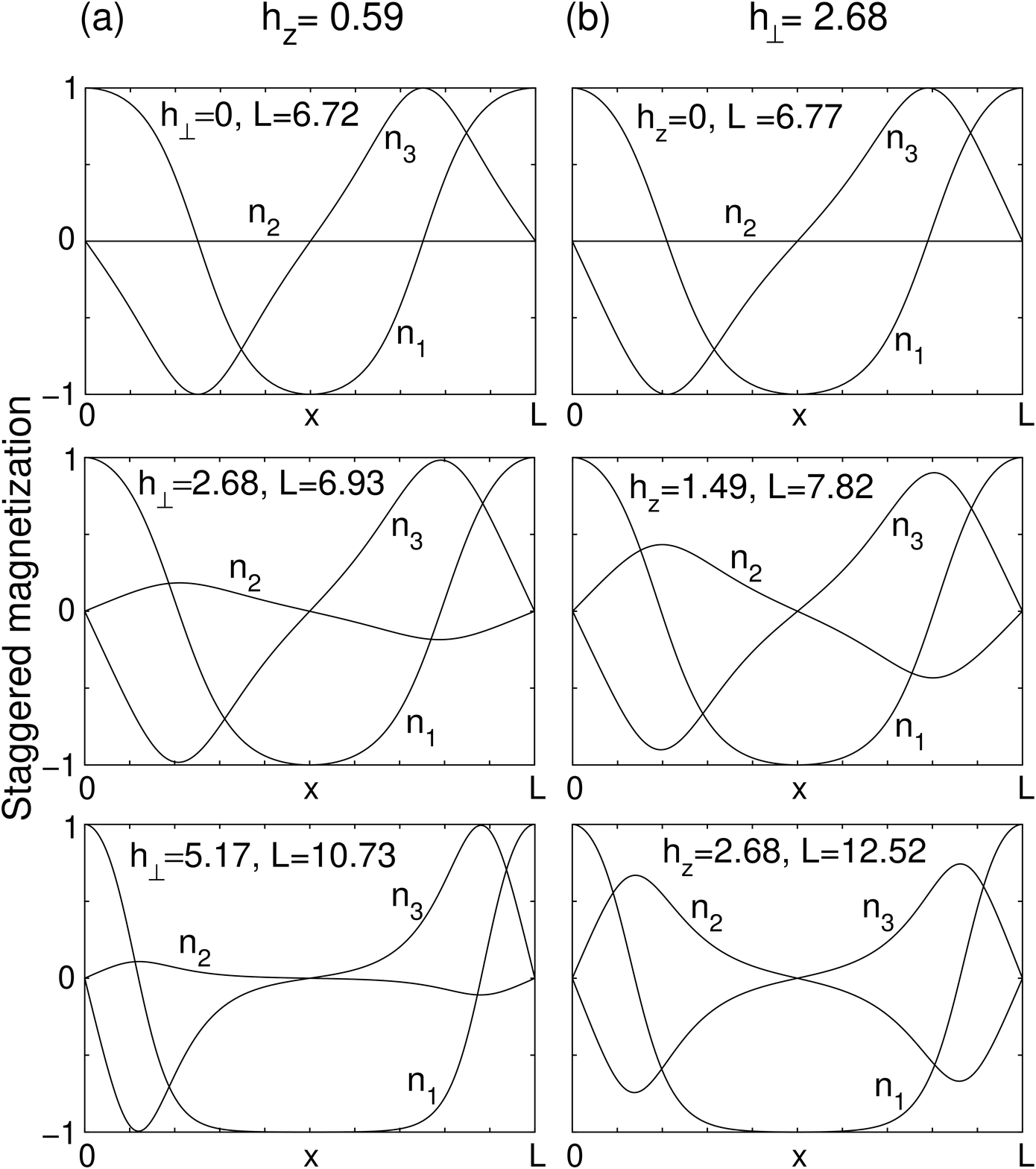} }
\caption{ Calculated evolution of spin configuration in the Antisymmetric phase
with the applied magnetic field. (a) Left panel: $h_{z}=0.59$ ($H_{z}=1$
T), $h_{\perp}$ increases. (b) Right panel: $h_{\perp}=2.68$ ($H_{\perp}=4.5$
T), $h_{z}$ increases. The bottom entries in both panels are applied
for points near the critical line between the Antisymmetric and the
Spin-flop phase. Notice an enhanced $n_{1}=-1$ domain in these entries.
The Antisymmetric spiral propagates strictly along $x$.}

\label{fig:5} 
\end{figure}

\begin{figure}
\resizebox{0.375\textwidth}{!}{\includegraphics{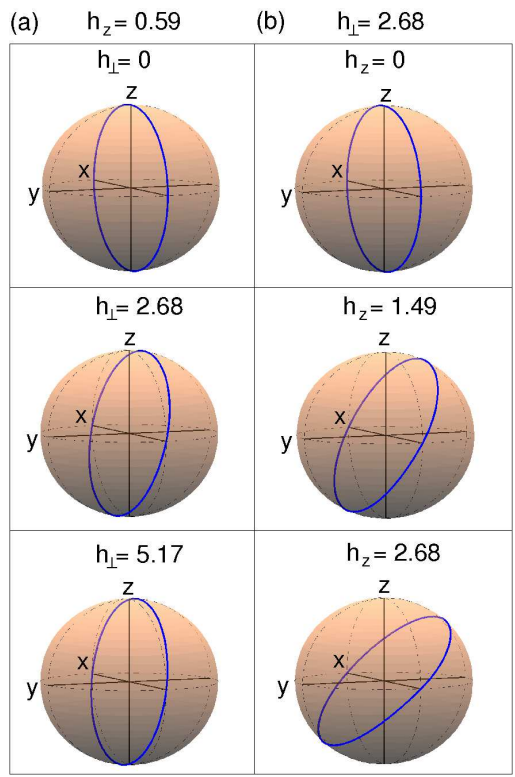} }
\caption{ (Color online) The same Antisymmetric spirals as in Fig.~5 but from
a different perspective. Blue lines on the sphere surface
are paths traced by the endpoint of ${\bf n}$ during one period
$L$.  }

\label{fig:6} 
\end{figure}

We begin our discussion with the case where $h_{z}<h_{c1}=1.01$ (or
$H_{z}<H_{c1}=1.7$ T) and consider the evolution of the system with
increasing $h_{\perp}$. Our results are displayed in Figs.~\ref{fig:5}(a) and ~\ref{fig:6}(a).
In the limit $h_{\perp}=0$, the spin configuration
that minimizes the energy is the flat spiral constructed in Sec.~\ref{sec:Field-parallel-to}.
Recall that this solution is degenerate with respect to rotations
around $c$, in agreement with the $U(1)$ symmetry given by Eq.~(\ref{eq:2.1}).
When $h_{\perp}\neq0$, the $U(1)$ symmetry is broken, and the energy
is minimized by a nonflat spin spiral propagating strictly along the $x$ axis.
The component of the staggered magnetization $n_{2}$ is now different
from zero and points in the direction of the transverse field $h_{\perp}$,
while its sign oscillates over the period $L=L(h_{z},h_{\perp})$
with the property $n_2(x)=-n_2(L-x)$. 
Because
of this characteristic behavior, we call this state the \emph{Antisymmetric
phase}. 
The path traced out by the staggered magnetization ${\bf n}$ during
one period $L$, shown in Fig.~\ref{fig:6}(a), looks relatively simple.
The spin rotates \emph{approximately} in a plane whose normal 
is tilted from the $y$ axis 
towards some new direction 
in the $yz$ plane.

The origin of the oscillating component $n_{2}$ can be understood
by a direct inspection of the Zeeman energy $\propto\left({\bf n}\cdot{\bf h}\right)^{2}$.
Its diagonal terms $n_{2}^{2}h_{\perp}^{2}$, $n_{3}^{2}h_{z}^{2}$
are always positive, but the off-diagonal contribution may become
\emph{negative} provided that $n_{2}$ adjusts so that its sign is
always opposite to the sign of $n_{3}$. 
But the projection of ${\bf n}$ onto the $xz$ plane rotates during the period $L$ thanks
to the chiral DM term $(n_1\partial_1n_3-n_3\partial_1n_1)$ in the potential $V$ of
Eq.~(\ref{eq:4.2}). Therefore, the sign of $n_{3}$ oscillates, and
$n_{2}$ also displays oscillatory behavior.

To fully describe the spin structure, all terms in the potential of
Eq.~(\ref{eq:4.2}) must be considered, but the main conclusion persists
-- the spiral minimizes its energy by developing $n_{2}\neq0$ \emph{along
the direction of the transverse field $h_{\perp}$}, and the sign of  $n_{2}$ oscillates
over the period $L$. As a result, the expectation value $\langle n_{2}n_{3}\rangle$
becomes negative ( $\langle n_{2}n_{3}\rangle<0$), while $\langle n_{2}\rangle=\langle n_{3}\rangle=0$,
as verified by a direct calculation.

We now briefly describe the role of the term $h_{\perp}d_{z}n_{1}$
in Eq.~(\ref{eq:4.2}). The importance of the latter contribution has
already been established in Sec.~\ref{sec:Field-perpendicular-to} during our analysis of the properties
of the flat spiral ($n_{2}=0$) in the presence of a field applied
strictly in the $xy$ plane ($h_{\perp}\neq0$, but $h_{z}=0$). The
scenario discussed in Sec.~\ref{sec:Field-perpendicular-to} is here mildly modified by the presence
of $h_{z}\neq0$ but it main features remain the same, as confirmed
by our numerical studies. The weak--ferromagnetic anisotropy $h_{\perp}d_{z}n_{1}$,
generated by the transverse field $h_{\perp}$ \emph{applied along
the y axis}, makes the spin orientations along the $\pm x$ axis energetically
nonequivalent. In the Antisymmetric state, the component of the staggered
magnetization that is perpendicular to the transverse field $h_{\perp}$
rotates in the $xz$ plane, and is thus directly affected by the weak--ferromagnetic
term $h_{\perp}d_{z}n_{1}$. In turn, the profile of the Antisymmetric
spiral is modified, and the expectation value of $n_{1}$ over the
period $L$ becomes nonzero and negative ($\langle n_{1}\rangle<0$)
in order to minimize $h_{\perp}d_{z}n_{1}$.

With increasing $h_{\perp}$, the spiral becomes significantly distorted,
and the $n_{1}\simeq-1$ orientation (domain) during the spin rotation
is greatly enhanced. This is apparent from the bottom entry of Fig.~\ref{fig:5}(a).
At the same time, the period $L$ of the spiral increases, and the
energy density of the Antisymmetric state begins to approach the energy
density of the uniform Spin-flop state ${\bf n}=(-1,0,0)$ from below.
At the critical value of the transverse field $h_{\perp}^{c}(h_{z})$,
the period of the spiral grows to infinity ($L\rightarrow\infty$),
and its energy density becomes equal to the energy density of the
Spin-flop state $w=\frac{1}{2}\left(1-2h_{\perp}d_{z}\right)$. This
numerically verified scenario is consistent with experiment \cite{17},
and is somewhat similar to that discussed in Sec.~\ref{sec:Field-perpendicular-to} for strictly
transverse fields. Above the critical line, only the uniform Spin-flop
states emerges from our numerical calculations, and the incommensurate
Antisymmetric spiral no longer exists. The boundary between the Antisymmetric
and the Spin-flop state is indicated by the solid line in Fig.~\ref{fig:4}.
We have verified that the Antisymmetric state displayed in the phase
diagram always carries lower energy density than the uniform Spin-flop
state.

Evolution of the spin structure with increasing
$h_{z}$, but fixed strength of the transverse field $h_{\perp}$,
is shown in Figs.~\ref{fig:5}(b) and \ref{fig:6}(b)
Our results, applied here for $h_{\perp}=2.68$ ($H_{\perp}=4.5$
T), are qualitatively similar to those in Figs.~\ref{fig:5}(a),~\ref{fig:6}(a).
In particular, the spiral again develops a nonzero oscillating
component $n_{2}\neq0$ along $h_{\perp}$, with zero expectation
value  $\langle n_{2} \rangle = 0$ over the period $L$.
Expectation value $\langle n_{2}n_{3}\rangle<0$
due to the off--diagonal anisotropy $n_{2}n_{3}h_{\perp}h_{z}$, whereas
$\langle n_{1}\rangle<0$ thanks to the 
weak--ferromagnetic term $h_{\perp}d_{z}n_{1}$. 
With increasing $h_{z}$, the period of the spiral $L$ increases
and presumably again diverges
($L\rightarrow\infty$) at the critical line. Above the critical line,
only the uniform Spin-flop states emerges from our numerical calculations,
and the incommensurate Antisymmetric spiral no longer exists.

We emphasize, that the
characteristic properties of the Antisymmetric state discussed in
the preceding paragraphs remain the same for \emph{any point} $h_{z}\neq0$,
$h_{\perp}\neq0$ below the solid line in Fig.~\ref{fig:4}. 
However, the scenario of the phase transition between the Antisymmetric
and the Spin-flop phase, discussed in connection with Fig.~\ref{fig:5},
is slightly modified for sufficiently weak $h_{\perp}$, near the
point $\Gamma$. Specifically, for $H_{\perp}$ below $\sim1$ T,
the energies of both states again become equal at the critical line,
but the period $L$ of the Antisymmetric spiral remains \emph{finite}
(albeit large). Above the critical line, our numerical minimization
still yields the solution in the form of the Antisymmetric spiral,
with however, the energy density higher than the energy density of
the Spin-flop state ${\bf n}=(-1,0,0)$. 
This should be contrasted
with behavior for large $h_{\perp}$, where the period of the Antisymmetric
spiral \emph{diverges} ($L\rightarrow\infty$) at the critical line;
and above the critical line only the uniform Spin-flop state exists.
Interestingly, our results seem to be again consistent with the experiment
\cite{17}.

\begin{figure}
\resizebox{0.48\textwidth}{!}{\includegraphics{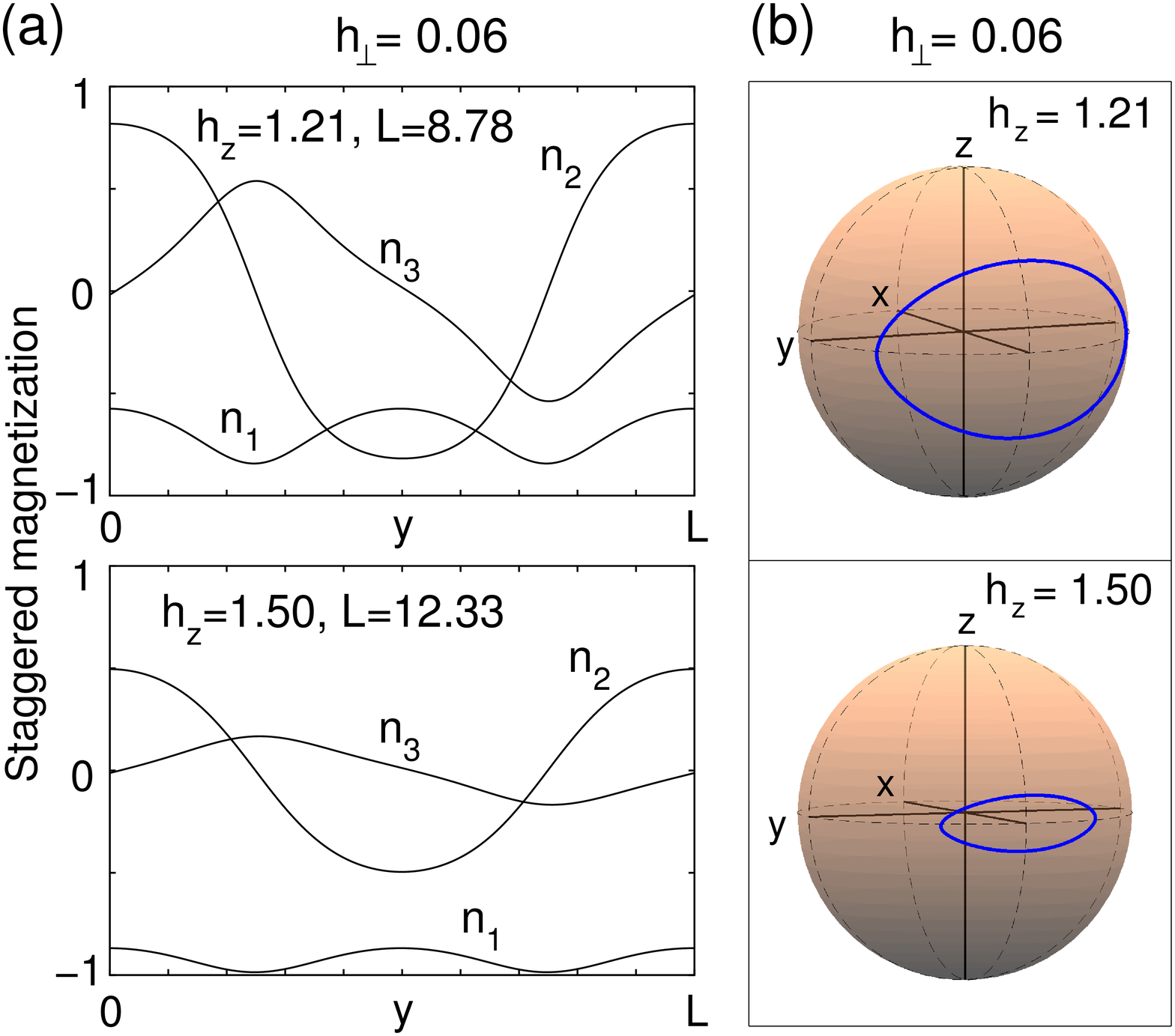} }
\caption{ (Color online)
(a) Examples of spin configurations in the Symmetric phase, calculated
for $h_{\perp}=0.06$ (or $0.1$ T) and the two values of $h_{z}$.
This phase exists in the narrow area of the phase diagram $1.01\lesssim h_z \lesssim \sqrt{3}$
($1.7$ T $<H_z< 2.9$ T)
and weak but nonzero $h_{\perp}$, $ 0 < h_{\perp} \lesssim 0.12$ ($0<H_{\perp}\lesssim 0.2$ T).
The Symmetric conical spiral propagates strictly along $y$ 
and nutates around the $-x$ axis.
Otherwise its properties are similar to its precursor phase, 
the Intermediate state of Sec.~IIIA.
(b) The same Symmetric spirals viewed 
from
a different perspective. Blue lines on the sphere surface
are paths traced by the endpoint of ${\bf n}$ during one period
$L$.}

\label{fig:7} 
\end{figure}

\vspace{5pt} 
We now focus on the area left from the point
$\Gamma$ in the phase diagram.
Our spin-wave analysis of the Spin-flop state
in the presence of arbitrary canted fields, given in Appendix B,
established that
the Spin-flop phase is locally unstable below the dashed line 
in Fig.~\ref{fig:4}.
Thus it cannot exist beyond the point $\Gamma$, where the energy density
of the Antisymmetric and the Spin-flop state become equal.
It is more or less clear, there is a new phase realized in some area
just below the dashed line, near the axis $h_{z}$ (near $h_{\perp}=0$).
Note that the dashed line starts from the point
$h_{z}=\sqrt{3}$ ($H_{z}=2.9$ T), which
is just the upper critical field $h_{c2}$ obtained in Ref.~24. 
For $h_{z}<h_{c2}$, the Spin-flop phase is locally unstable, 
and the Intermediate phase is realized in the region $h_{c1}<h_{z}<h_{c2}$. 
It is 
natural
to expect that the Intermediate phase  
survives
in some form also in the
presence of a weak transverse field $h_{\perp}\neq0$.

Our calculations confirm this expectation.
When $h_{\perp}\neq 0$, a 
\emph{Symmetric phase} emerges as the ground state 
in the region between the dashed and the solid line in Fig.~\ref{fig:4}.
This phase acquires its name because $n_1(y)=n_1(L/2+y)$.
Two examples are illustrated in Fig.~\ref{fig:7}.
The Symmetric phase can be described as an 
antiferromagnetic conical spiral that
propagates \emph{strictly along y}, but nutates around the $-x$ axis.
Importantly, the component $n_{1}$, \emph{perpendicular} to
the transverse field $h_{\perp}$, is nonzero ($n_{1}\neq 0$) and 
always negative ($\langle n_{1}\rangle < 0$). 
All these features agree with the experiment \cite{17}.

\begin{figure}
\resizebox{0.4\textwidth}{!}{\includegraphics{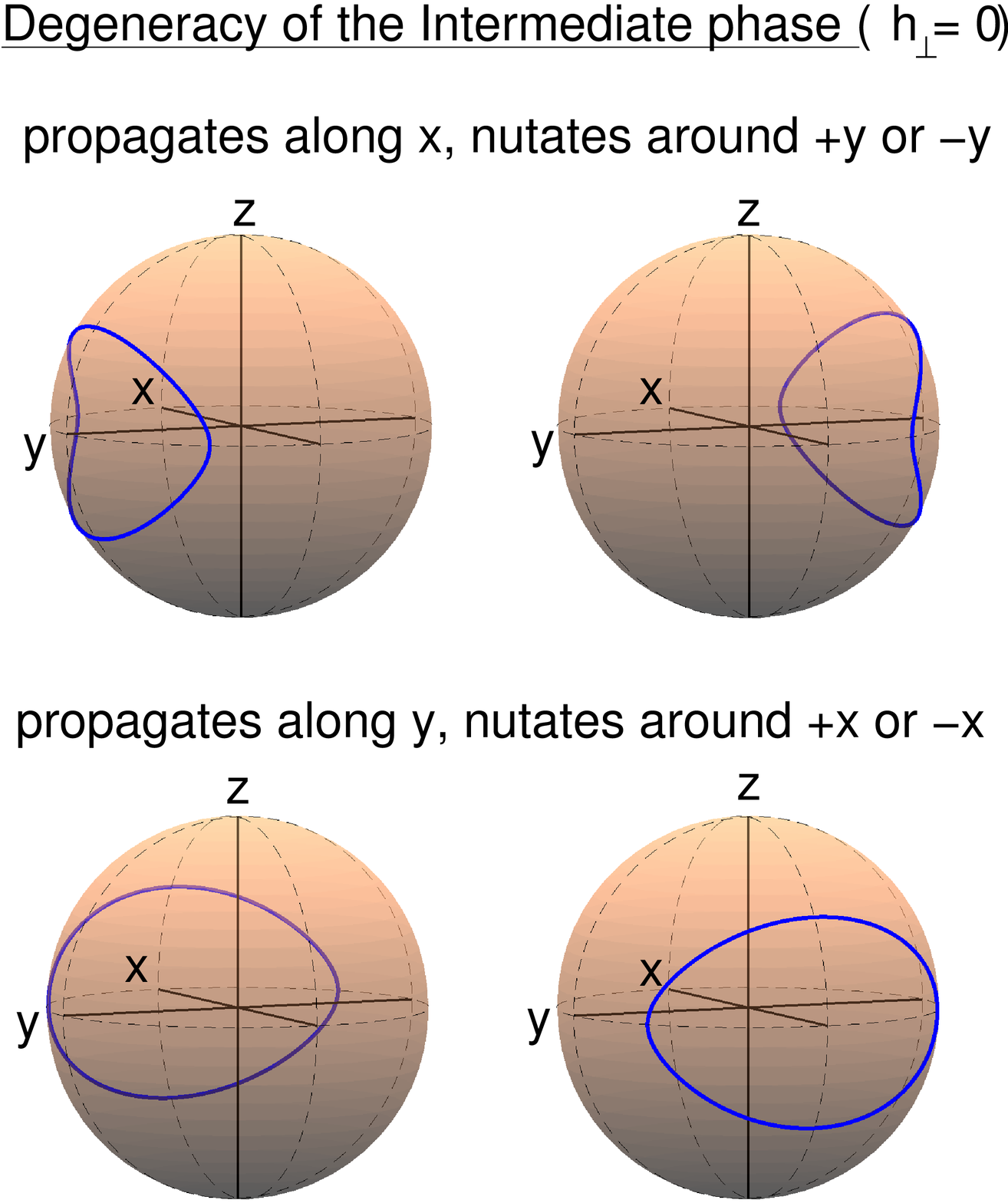}} 
\caption{ (Color online) Illustration of the four degenerate
states in the Intermediate phase ($h_{\perp}=0$) discussed in the text,
calculated here for $h_{z}$=1.21. 
Thick lines on the sphere indicate paths
traced out by the endpoints of the staggered magnetization during
one period $L$. The base of the staggered magnetization is placed
at the center of the sphere.
A nonzero transverse component $h_{\perp}\neq 0$,
applied along $+y$, breaks the above degeneracy and
favors the spiral propagating along $y$ and nutating around $-x$ axis,
which becomes the precursor of the Symmetric spiral.
}

\label{fig:8} 
\end{figure}
\begin{figure}
\resizebox{0.48\textwidth}{!}{\includegraphics{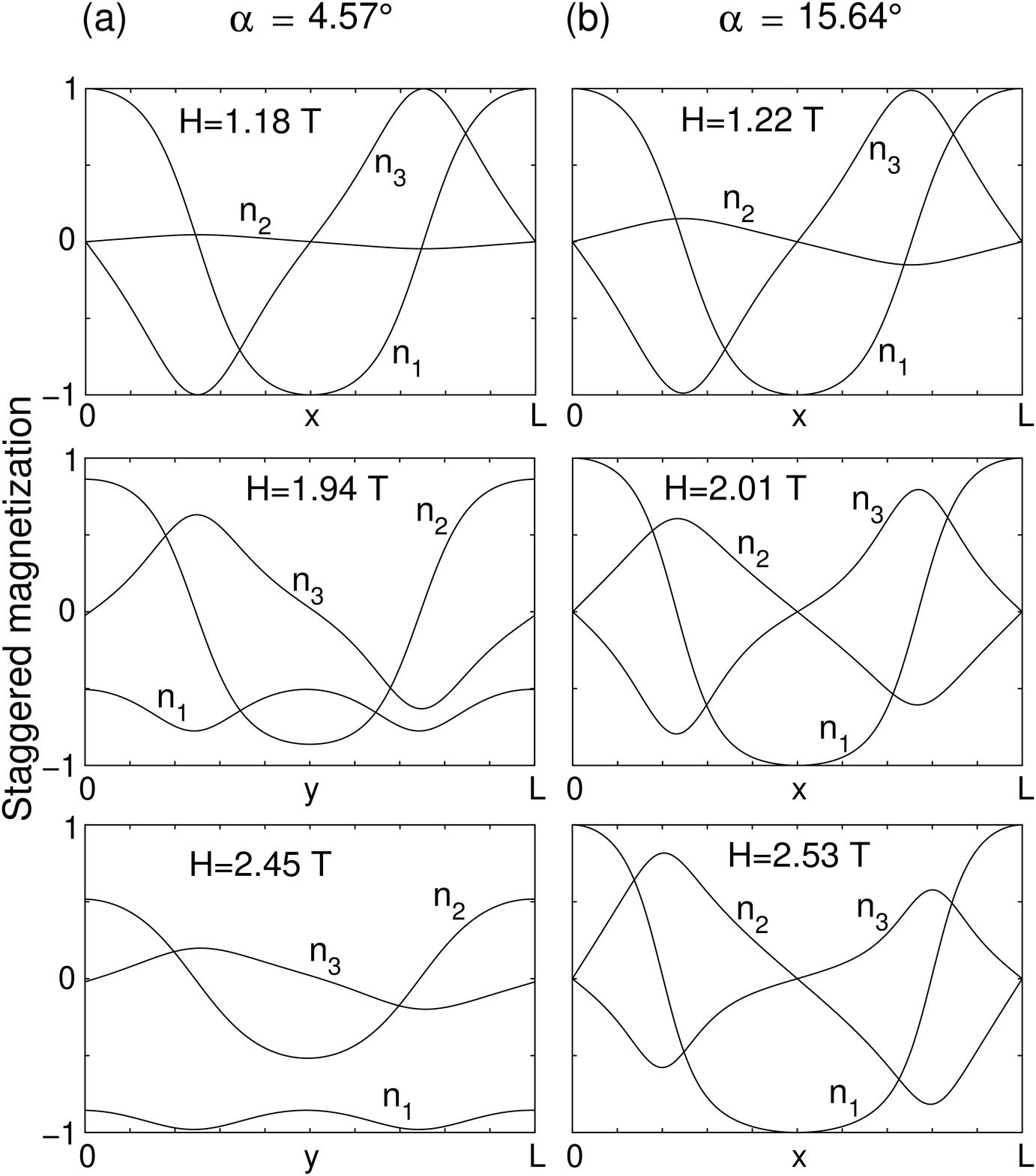} } 
\caption{
Calculated spin configurations in a magnetic field
of varying strength $H$, applied at an angle $\alpha$ with respect to the $c$ axis.
The two values of  $\alpha$ are roughly equal to $5^{\circ}$ and  $15^{\circ}$ used
in neutron scattering measurements of Ref.~29.
(a) $\alpha=4.57^{\circ}$. The top entry corresponds to the Antisymmetric phase, 
and the spiral propagates along $x$. 
The middle and the bottom entry display the Symmetric phase, where
the spiral propagates along $y$. Phase transition between
the Antisymmetric and the Symmetric phase is accompanied by sudden
$\pi/2$ rotation of propagation direction.
(b) $\alpha=15.64^{\circ}$. All entries show the Antisymmetric phase, 
with propagation direction along $x$.}

\label{fig:9} 
\end{figure}

The Symmetric phase develops from its predecessor, the Intermediate phase
-- the conical antiferromagnetic spiral -- 
discussed in \ref{sec:Field-parallel-to}
for fields strictly parallel to $c$.
Recall that the Intermediate phase  
obeys the U(1) symmetry described by Eq.~(\ref{eq:2.1}).
In practice, the U(1) symmetry is broken by an additional tetragonal anisotropy induced by
discreteness effects \cite{6,15}.  
Thus, in the absence of transverse fields, there exist four degenerate 
states, shown in Fig.~\ref{fig:9}:
the conical spiral propagates along $x$ and nutates around the $\pm y$ axis; 
or it propagates along $y$ but nutates around the $\pm x$ axis.

This degeneracy is broken when $h_{\perp}\neq 0$.
To illustrate this point, consider for a moment 
the Intermediate spiral with the profile ${\bf n}$, calculated
in the absence of a transverse field.
Inserting
this solution in the potential $V$ of Eq.~(\ref{eq:4.2}),
applied with $h_{\perp}\neq 0$, yields the additional
corrections to the energy given by 
$\frac{1}{2}h_{\perp}^{2}\langle n_{2}^{2}\rangle+h_{\perp}d_{z}\langle n_{1}\rangle$.
The first correction, quadratic in $h_{\perp}$, originates in the Zeeman energy
$\propto \left({\bf n}\cdot{\bf h}\right)^2$.
Note that the off--diagonal Zeeman term $h_{\perp}h_{z}\langle n_{2}n_{3}\rangle$
does not contribute, because
the expectation value $\langle n_{2}n_{3}\rangle$ in the Intermediate phase
vanishes for any degenerate state. 
The second correction, linear in $h_{\perp}$, is due to the weak--ferromagnetic
anisotropy $d_{z}({\bf h}\times{\bf e}_{3})\cdot{\bf n}$. 
This 
linear contribution dominates for small transverse field, 
and favors 
the particular degenerate state; namely,   
the conical spiral propagating along $y$ and nutating around the $-x$ axis,
with $\langle n_{1}\rangle <1$.
Numerical work confirms that the above qualitative argument is correct despite
the simplifying assumption that neglects the changes in the staggered
magnetization induced by $h_{\perp}$. 
The actual profile of the Symmetric spiral ${\bf n}$ and its period
$L$ are both mildly modified by $h_{\perp}\neq0$. Otherwise its properties
are similar to the Intermediate phase.

The Symmetric phase emerges in canted magnetic fields applied nearly
parallel to the $c$ axis, when $h_{z}\gtrsim h_{c1}$. 
It is the stationary point of the energy functional
with the lowest energy density in the area
between the dashed line and the solid line of Fig.~\ref{fig:4}.
With increasing
$h_{z}$ 
the period $L$ increases, and 
the magnitude of $n_{1}$ becomes larger and larger, until
at the dashed line $n_{1}\rightarrow-1$ and the solution becomes
the Spin-flop state ${\bf n}=(-1,0,0)$. 
This behavior, apparent also from Fig.~\ref{fig:7}, is 
virtually identical to the Intermediate phase. 
Our results generally agree
with experimental findings \cite{16,17}.
Evolution of the spin structure with increasing $h_{\perp}$, but
fixed strength of the longitudinal component $h_z$ is rather mild.
The period $L$ slightly decreases with $h_{\perp}$, whereas the
magnitude of $n_1$ moderately increases due to the weak--ferromagnetic
energy  $d_{z}({\bf h}\times{\bf e}_{3})\cdot{\bf n}$.
Importantly, at the critical solid line 
the energy density of the Symmetric phase becomes
equal to the energy density of the Antisymmetric phase.
This 
happens at $h_{\perp} \sim 0.12$ or 0.2 T.
For stronger $h_{\perp}$, the Antisymmetric state
emerges as the true ground state, with the energy density lower than the Symmetric spiral.
The corresponding phase transition is first order, and is further discussed 
in the following paragraphs.
\\
\\
{\bf Comparison with neutron diffraction.}
Experimental data were obtained from measurements with
a magnetic field of varying strength $H$ 
applied at an angle ${\alpha}$ with respect to the $c$ axis\cite{16,17}.
See the straight (green) lines in Fig.~\ref{fig:4}.
The in-plane component of the field $H_{\perp}=H\sin{\alpha}$ 
was directed along the $y$ axis, or the (-1,1,0) axis using the notation of Refs.~[28,29].  
The alternative choice, $H_{\perp}{\parallel}$ (1,0,0), yielded
equivalent results, and is thus ignored in the following discussion.
We concentrate on the data for 
${\alpha}\sim 5^{\circ}$ and $15^{\circ}$, analyzed in detail in Ref.~[29].
We used similar angles $4.57^{\circ}$ and $15.64 ^{\circ}$ in our calculations.

\begin{figure}
\resizebox{0.48\textwidth}{!}{\includegraphics{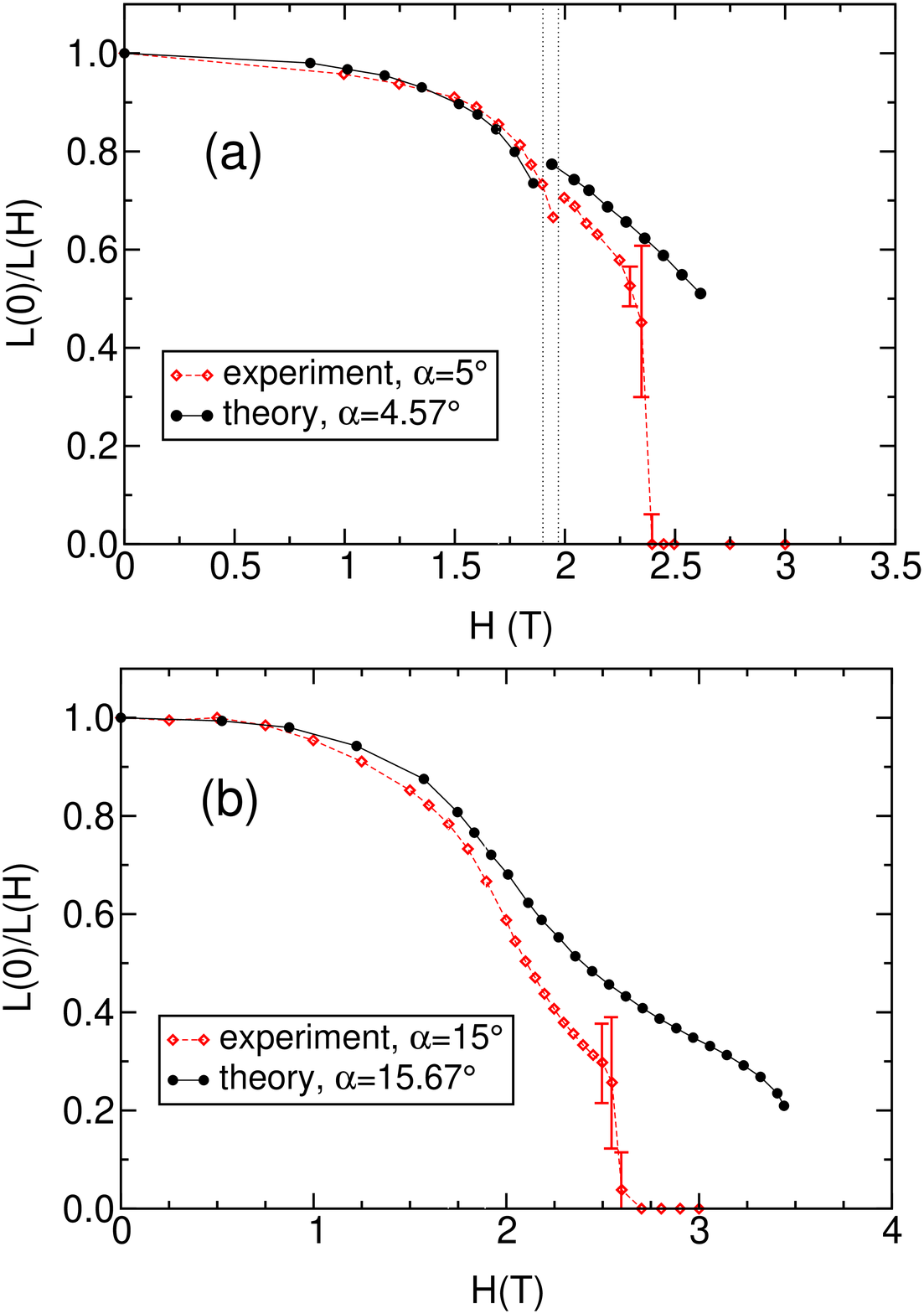} } 
\caption{(Color online) 
Evolution of the incommensurability parameter $L(0)/L(H)$
with 
the strength of 
the magnetic field applied at an angle $\alpha$
with respect to the $c$ axis. 
Comparison of $T=0$ teoretical predictions with experiment. 
Neutron scattering data and error bars, taken at $T=1.65$ K,
were extracted from Fig.~10, Ref.~29 and adjusted
to fit our conventions. 
(a) $\alpha\approx 5^{\circ}$. 
The two dotted lines mark the location of the critical field
$H_{c1}(\alpha)$ determined by theory ($1.9$ T) 
and experiment ($1.97$ T).
A discontinuous jump to a larger value at $H_{c1}$ 
corresponds to the phase transition between the Antisymmetric
and the Symmetric phase.
(b) $\alpha\approx 15^{\circ}$.
}

\label{fig:10} 
\end{figure}

\begin{figure}
\resizebox{!}{.65\textheight}{\includegraphics{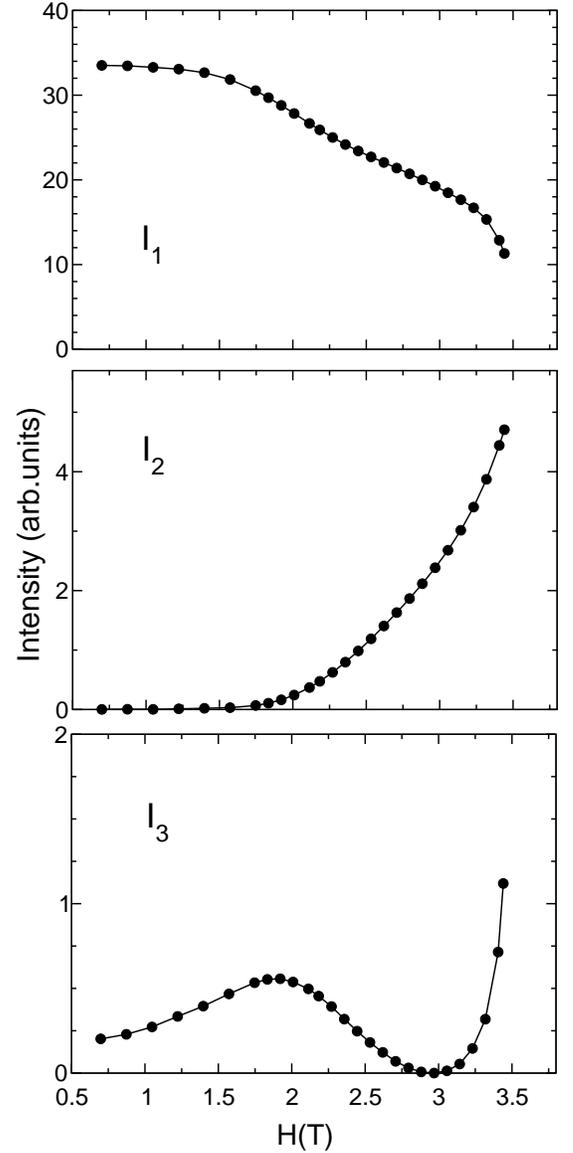} } 
\caption{Theoretical field dependence for the intensities of the 1st, 2nd and
the 3rd Fourier harmonics calculated from the $n_{1}$ and $n_{3}$
components of the staggered magnetization. The magnetic field
was $H$ was applied at an angle $\alpha=15.64^{\circ}$
with respect to the $c$ axis and is roughly
equal to 15$^{\circ}$ used in the experimental Fig.~8, Ref.~29.
We adopt SI units to facilitate comparison with the experiment. }

\label{fig:11} 
\end{figure}

We first discuss the case ${\alpha} \sim 5^{\circ}$ 
For $H < H_{c1}(\alpha)= 1.95$ T, the experiment
observed an incommensurate structure that propagates along $y$,
while its spin rotates in the $xz$ plane, perpendicular to $H_{\perp}$.
This structure is a cycloid for weak $H$, that distorts to a soliton
lattice for stronger fields.  
At the critical field $H_{c1}$, the propagation direction suddenly rotates 
exactly by $\pi/2$, from the $y$ to the $x$ axis, and the structure becomes
an antiferromagnetic cone\cite{16,17}.
This antiferromagnetic cone phase (or the double-$k$ phase) propagates
along $y$, parallel to $H_{\perp}$. Its ``incommensurate'' spin component rotates
in the $yz$ plane, while a ``commensurate'' component is perpendicular to both
the $c$ axis and the transverse field, as shown in Fig.~13(a) in Ref.~29.
Finally, the IC transition is observed at $H_{c1}\approx 2.4$ T.

These experimental findings are consistent with our results. 
For $H < H_{c1}$ (the top entry in Fig.~\ref{fig:9}(a)), 
theory predicts the Antisymmetric phase that propagates
strictly along $x$, perpendicular to $H_{\perp}$. 
Importantly, the component of the staggered magnetization $n_2$, oscillating along
$H_{\perp}$, is for $\alpha \sim 5^{\circ}$ rather small, and the structure
resembles the flat spiral. The Antisymmetric phase can be identified
with the cycloid and/or the soliton lattice of Ref.~[29].
Above  $H_{c1}(\alpha)$, predicted at $\sim 1.9$ T, the Symmetric phase --
a conical spiral propagating along $y$ but nutating along the $-x$ axis --
emerges, see the middle and the bottom entry in Fig.~\ref{fig:9}(a).
This agrees with the spin structure proposed in Ref.~[29], and the Symmetric phase should be 
identified
with the antiferromagnetic cone phase.
The IC transition occurs at $H_{c2}(\alpha)\sim 2.7$ T.
Note that the Antisymmetric spiral propagates strictly along the $x$ axis, while the Symmetric
propagates along $y$. Therefore, the phase transition at  $H_{c1}$
is accompanied by a sudden rotation of the spiral propagation direction 
exactly by ${\pi}/{2}$, as in the experiment. 
This sudden ${\pi}/{2}$ rotation, highlighted as a noteworthy feature of recent experiments,
has been explained by our previous analysis earlier in this section.
Specifically, the presence of $H_{\perp}\neq 0$ breaks the $U(1)$ symmetry described by Eq.~(\ref{eq:2.1}),
and selects a particular spin orientation and spiral propagation direction in each phase. 
Thus, in the Antisymmetric phase, propagation along $x$ is required to minimize the Zeeman energy;
while the Symmetric phase must propagate along $y$ in order
to minimize the dominant weak--ferromagnetic contribution $d_{z}({\bf h}\times{\bf e}_{3})\cdot{\bf n}$.
In this respect we note that the sudden ${\pi}/{2}$ rotation should only occur when $H_{\perp}\neq 0$,
and is not expected for $\alpha=0$ or fields applied strictly along the $c$ axis.
However, a perfect alignment of the applied field with the $c$ axis
is impossible to achieve in practice. Therefore the results of Ref.~[28]
reported for $\alpha=0$ and ``a sample with an almost perfect alignment'' (misalignment less than $0.5^{\circ}$)
should be interpreted from the perspective of our previous comment.

We now discuss the evolution of the incommensurability parameter
$L(0)/L(H)$, shown in Fig.~\ref{fig:10}(a).
Note a discontinuous jump of the incommensurability parameter to higher value at
the critical field $H_{c1}$, considered as another characteristic feature of
the phase transition to the antiferromagnetic cone phase\cite{17}.
We emphasize a remarkable agreement of our theory with the experimental data;
the incommensurability parameter increases by  $\sim5\%$ (theory) or
 $\sim6\%$ (experiment). The calculated value of $H_{c1} \sim 1.89$ T also agrees well
with $\sim 1.97$ T extracted from the experimental data. Note that the latter
value, obtained from our analysis of the experiment, 
differs from 1.95 T quoted in Ref.~[29]. 
The calculated critical field $H_{c2}\sim 2.7$ T is slightly larger than
the observed $\sim 2.4$ T.
However, the overall agreement is good. 
One should keep in mind that we compare the $T=0$ calculations with
the data taken at relatively high temperature  $T=1.65$ K$\sim 0.5~T_{N}$.
Actually our theoretical data are shown only for field values up to
$2.6$ T $< H_{c2}$. 
This is because of numerical difficulties that occur
as the period $L$ rapidly quickly grows near $H_{c2}$.
The corresponding average energy density, which is a function of $L$, then
displays a shallow minimum that makes it hard to determine
the precise value of $L$.
Nevertheless, our calculations indicate
a continuous IC transition with $n_1\rightarrow-1$, but \emph{finite} (albeit large)
$L$ in the limit $H\rightarrow H_{c2}$.

We now discuss the case when the field is applied 
at ``large''
angle $\alpha\sim 15^{\circ}$ with respect
to the $c$ axis. 
In this case, no reorientation of the spiral propagation direction 
(which is along $x$) was observed in the experiment \cite{17}.
For $H\gtrsim 1.7$ T, the proposed spin structure was
described as clearly non-sinusoidal, non-planar ``complexly distorted incommensurate phase'', 
whose detailed structure, however, remained unresolved.
This ``distorted incommensurate phase'' was characterized
by the smooth appearance of higher order harmonics seen by neutron diffraction, both odd and even.
The measured dependence of the incommensurate parameter on $H$  displays 
a characteristic shape, concave for
weak field, and convex when 
$H\gtrsim1.7$ T.
Finally, the IC transition is observed at $\sim 2.6$ T.
All these features are consistent with
our $T=0$ calculations, which predict the Antisymmetric phase with
oscillating $n_{2}\parallel H_{\perp}$, propagating
along $x$ for all field strengths until the IC transition 
at $\approx3.45$ T. 
Predicted critical field is somewhat larger than that observed in the
experiment, but is not terribly inconsistent with the measured value $2.6$
T quoted in Ref.~29,
especially in view of our previous comments.
The $n_2$ component is small for weak fields, but its magnitude
quickly increases with $H$, as apparent from 
Fig.~\ref{fig:9}(b). 
For stronger fields, the calculated structure becomes clearly non-sinusoidal, 
non-planar and can
be identified with the ``distorted incommensurate structure'' 
of Ref.~29. 
Importantly, the Fourier transform of the staggered magnetization
provides evidence for higher harmonics, both odd and even.

Our $T=0$ theoretical results for the field dependence of the \emph{intensities}
of the 1st, 2nd and 3rd Fourier components of the staggered magnetization
${\bf n}(x)$ are presented in  Fig.~\ref{fig:11}.
The intensities are calculated from the $n_{1}$, $n_{3}$
components. This is because neutron scattering sees only the components
perpendicular to momentum transfer (which is parallel to $n_{2}$).
Our results are related to the experimental data in Fig.~8(b)
of Ref.~29. 
The 1st harmonic displays the typical shape seen in the experiment; first
very mild, almost linear decrease folowed by a convex shape for $H\gtrsim 1.7$ T
that becomes concave near the IC critical field. 
The difference with experiment thus lies mainly in somewhat larger theoretical value
of the IC critical field, as mentioned already in the previous paragraph. 
In agreement with the observation, higher harmonics smoothly appear above 1.7
T - 2. The intensity of 2nd harmonics linearly increases with
the field, as in the experiment. 
Similarly, the 3rd harmonics first
increases, then shows a shallow dip and increases again near the
IC critical field. 
This characteristic behavior is exactly what was observed in the experiment\cite{17}.
On the other hand, our results show rapid increase of
both higher harmonics as the field approaches the critical value,
whereas the experimental data show smoothing at the IC transition.
This can be perhaps due to finite temperature. Overall agreement with
experiment is however fairly good. 

Finally we discuss the field dependence
of the incommensurability parameter shown Fig.~\ref{fig:10}(b)
In agreement with experiment, the curve shows no discontinuity until the IC transition.
The shape is concave for
weak field strengths, but becomes convex for 
$H\gtrsim1.7$ T. 
This corresponds to the emergence of higher harmonics and is again in agreement
with the experiment, with minor
discrepancy in the value of the IC critical field.
The nature of the observed IC transition as deduced from the measured incommensurability
parameter remains unclear; the corresponding wording
in p.~9, Ref.~[29] suggests that the data are consistent with a discontinuous transition.
However, the related discussion in p.~8 of the same reference states that due to the
smallness of the measured parameter near the critical field ``no reliable conclusion
can be drawn''  whether the data continuously diverge or show a finite jump.
In any case, for $\alpha=15^{\circ}$ indicates a discontinuous IC transition,
which becomes continuous for larger angles. This point has already been briefly discussed 
in the discussion of the Antisymmetric spiral. 
\\
\\
{\bf Comparison with magnetic susceptibility measurements.}
Neutron scattering studies confirmed the existence of
the double-$k$ phase and proved useful for examination of its properties.
However, they were limited to only few values $\alpha=0^{\circ}$, $5^{\circ}$,
$15^{\circ}$ and $30^{\circ}$.
Thus the exact boundaries of the
double-$k$ phase (or the Symmetric phase) remains an open question.
For example, our theory predicts that the latter phase exists for
$\alpha \lesssim 6^{\circ}$, which is just slightly above the experimentally studied case $5^{\circ}$.
Additional measurements in the region  $5^{\circ} < \alpha < 10^{\circ}$ may help to clarify this issue.

The phase diagram was further explored
by  complementary magnetic susceptibility measurements.
Peaks in the experimental data, taken at $T = 1.8$ K,  yielded the two critical ``lines''
marked in the phase diagram of Fig.~\ref{fig:4} by crosses
and diamonds.
Near the $c$ axis($\alpha=0^{\circ}$ and $5^{\circ}$) a single sharp peak in the
data was interpreted as the transition to the double-$k$ phase at $H_{c1}$ . 
The results were practically identical with neutron scattering studies, 
as apparent from the overlap of the first two crosses with neutron 
diffraction data. The IC transition at $H_{c2}$
is featurless in magnetic susceptibility.
 
However, for $\alpha \geq 10^{\circ}$ a single peak splits into two.
The lower peak (crosses) is interpreted as a crossover to a ``distorted
incommensurate structure'' seen in magnetic diffraction.
The nature of this ``distorted structure'' has already been discussed in previous
paragraphs. The lower peak broadens with increasing $\alpha$, and completely disappears at
$\alpha \sim 45^{\circ}$.
The sharp upper peak (diamonds) is clearly seen until
$\alpha \sim 90^{\circ}$ corresponds to the IC phase transition between
the Antisymmetric and the Spin-flop phase.
The agreement between our $T=0$ theory and the experiment is
almost perfect for $\alpha \sim 90^{\circ}$ or the fields applied strictly in the $xy$ plane.
This is not surprising, because the measured critical field 9 T were actually used
as an input value in our theoretical estimate of the out-of-plane DM anisotropy $d_z$.
The slight discrepancy in the critical field for strictly transverse field, 
seen in the phase diagram of Fig.~\ref{fig:4}, is simply due to the fact that we adopted 
the rounded value $d_z=0.06$ that differs by $\sim 10^{-3}$ from the exact result.
For canted fields,  our theory predicts somewhat larger critical fields than
those measured in the experiment.  
However, the overal agreement is satisfactory, and
discrepancies in the values of the critical fields are typically $\sim$10--20\%

We end this section with two comments:
\begin{itemize}
\setlength{\itemsep}{1pt}%
\item Numerical work confirms that the existence of $d_{z}\neq0$ is not
crucial for the appearance of the \emph{double-k structure} and/or
sudden ${\pi}/2$ rotations observed in experiment. It is, however,
important to provide quantitative agreement with experiment. In particular,
for $d_{z}=0$ and fields nearly parallel to $c$, the intermediate
phase would first appear at $h_{c1}$ in the form of a nonflat spiral
propagating strictly along $x$ but nutating around the $y$ axis,
\emph{without} reorientation of the spin propagation direction. A
sudden ${\pi}/2$ rotation of the propagation direction occurs later,
at yet another critical field $h_{rotation}\approx1.20$, above which
the minimum energy state becomes the Symmetric spiral propagating
along $y$ but nutating around the $x$ axis. In the absence of the
weak--ferromagnetic energy $d_{z}({\bf h}\times{\bf e}_{3})\cdot{\bf n}$,
an explanation of sudden reorientation requires a detailed analysis
of the of the energy term $\propto\left({\bf n}\cdot{\bf h}\right)^{2}$. 
\item We assumed that the transverse component of the field $h_{\perp}$
points strictly along the $y$ axis. Our results, however, are not
restricted to this special case. For example, assume that $h_{\perp}$
points in an arbitrary direction in the $xy$ plane, which is obtained
by a clockwise rotation of the $y$ axis with angle $\psi$. Then
the staggered magnetization ${\bf n}$ for any state calculated earlier
in this section must be also rotated clockwise with the angle $\psi$
around the $c$ axis, while the original direction of spin propagation
must be rotated counter-clockwise, with the angle $-\psi$. All other
results remain unchanged. 
\end{itemize}

\section{Conclusion}

\label{sec:Conclusion} We have presented a rather complete theoretical study
of $T=0$ phase transitions in canted fields of arbitrary strength
and direction. We calculated the complete phase diagram and identified
the symmetries of states in a number of different regions. For the
fields applied nearly parallel to the $c$ axis, we confirmed the
existence and stability of the Intermediate phase that mediates
the incommensurate-commensure transition and analyzed its properties.
We identify this phase with an experimentally observed 
\emph{double-k structure}. By analyzing data on fields applied 
perpendicular to the
$c$ axis, we determine an out-of-plane anisotropy parameter $d_{z}$
needed to complete quantitative comparison with experiment. Finally,
our model accounts for sudden ${\pi}/2$ rotations that have been
highlighted as a noteworthy feature of recent experiments. 

The work reported in this paper results from a long-standing theoretical
investigation of spiral magnetic structures in Dzyaloshinskii-Moryia antiferromagnets.
The theoretical framework involves a number of approximations: the
replacement of quantum-mechanical by classical variables, ignoring inter-layer couplings 
and the replacement of discrete spins by continuous fields in a model Lagrangean. 

Nevertheless, detailed agreement with experiment\cite{16,17} is now
so extensive that the applicability of this model to systems such
as $\mathrm{Ba_{2}CuGe_{2}O_{7}}$ may now be established. The only
remaining discrepancies lie in the particular magnetic field values
at which transitions between magnetic states take place.
These discrepancies are on the order of 10--20\%, which is not much beyond
experimental uncertainty. 
The discrepancies can also be partly attributed to the fact 
that the experimental data were taken at relatively high temperature 
$\sim 0.5~ T_N$.

\acknowledgments This work was partially supported by ESF Research
Networking Programme POLATOM. J.C. gratefully acknowledges the support
by the Slovak Research and Development Agency under the contract No.
APVV-0027-11. M.M. acknowledges partial support also from the U.S.
National Science Foundation through DMR1002428.
The work of N.P. has been co-financed by the European Union (European Social Fund, ESF) 
and Greek national funds through the operational program Education and Lifelong Learning of 
the National Strategic Reference Framework (NSRF) under 
"Funding of proposals that have received a positive 
evaluation in the 3rd and 4th Call of ERC Grant Schemes".

\appendix
\section{Magnon spectrum for $h{\parallel}c$}
\label{sec:magnon_spectrum_h_parallel_c}

Here we calculate the magnon spectrum of the intermediate state from
Sec.~\ref{sec:Field-parallel-to}.
We first introduce new fields according to
\begin{eqnarray}
\Theta(x,y,t) & = & \theta(x)-g(x,y,t)\nonumber \\
\Phi(x,y,t) & = & \phi(x)+f(x,y,t)/\sin\theta(x)
\label{eq:2.14}
\end{eqnarray}
where $\theta$ and $\phi$ are solutions for the intermediate state
found previously in Sec.~\ref{sec:Field-parallel-to}, and
$f$ and $g$ account for small fluctuations. 
The new fields  (\ref{eq:2.14})
are introduced in the complete Lagrangian
given by  Eq.~(\ref{eq:1.4}) applied with
${\bf h}=(h,0,0)$ which is then expanded to second
order in $f$ and $g$.

The final result for the linearized equations of motions is
\begin{eqnarray}
(\partial_{1}^{2}+\partial_{2}^{2}-\partial_{0}^{2})f & = & U_{11}f+U_{12}g+A\partial_{1}g+B\partial_{2}g+C\partial_{0}g\nonumber \\
(\partial_{1}^{2}+\partial_{2}^{2}-\partial_{0}^{2})g & = & U_{22}g+U_{21}f-A\partial_{1}f-B\partial_{2}f-C\partial_{0}f\nonumber \\
\label{eq:2.15}
\end{eqnarray}
where all functions except $f$ and $g$ are functions only of $x$
and are given by
{
\allowdisplaybreaks
\begin{eqnarray}
U_{11} & = & -\left(\partial_{1}\theta\right)^{2}+\cos^{2}{\theta}\left(\left(\partial_{1}\phi\right)^{2}\,-2\,\partial_{1}\phi\right)\nonumber \\
 &  & +{\gamma}^{2}\,\left(\cos^{2}{\phi}\,\cos^{2}{\theta}-2\,\,\cos^{2}{\phi}+1\right)\nonumber \\
\nonumber \\
U_{12} & = & -\frac{{\left(2\,\partial_{1}\phi-2\right)\,\partial_{1}\theta}}{\sin{\theta}}\nonumber \\
\nonumber \\
U_{21} & = & \frac{{\left(2\,\partial_{1}\phi-2\right)\,\cos^{2}{\theta}\,\partial_{1}\theta
+2\,{\gamma}^{2}\,\cos{\phi}\,\sin{\phi}\,\cos{\theta}\,\sin{\theta}}}{{\sin{\theta}}}\nonumber \\
\nonumber \\
U_{22} & = & -\left(\left(\partial_{1}\phi\right)^{2}-2\,\partial_{1}\phi+{\gamma}^{2}\,\cos^{2}{\phi}\right)\,\left(2\,\sin^{2}{\theta}-1\right)\nonumber \\
\nonumber \\
A & = & \left(2\,\partial_{1}\phi-2\right)\,\cos{\theta}\nonumber \\
\nonumber \\
B & = & -2\,\sin{\phi}\,\sin{\theta}\nonumber \\
\nonumber \\
C & = & -2\,{h}\,\cos{\phi}\,\sin{\theta}.\label{eq:2.16}
\end{eqnarray}
}
We have verified that for the flat spiral ($\theta=\pi/2,$ $\partial_{1}\phi=\sqrt{\delta^{2}+\gamma^{2}\cos^{2}\phi}$)
these expressions reduce to those previously obtained for the magnon
spectrum of the flat spiral in Eq.~(5.2) of Ref.~24, but
with $\phi\longleftrightarrow\theta$. 
Note that the two linear equations for $f$ and $g$ are coupled as long as the magnetic field $h$
is different from zero or spin wave propagation deviates from the $x$ axis.

\begin{figure}
\resizebox{0.47\textwidth}{!}{\includegraphics{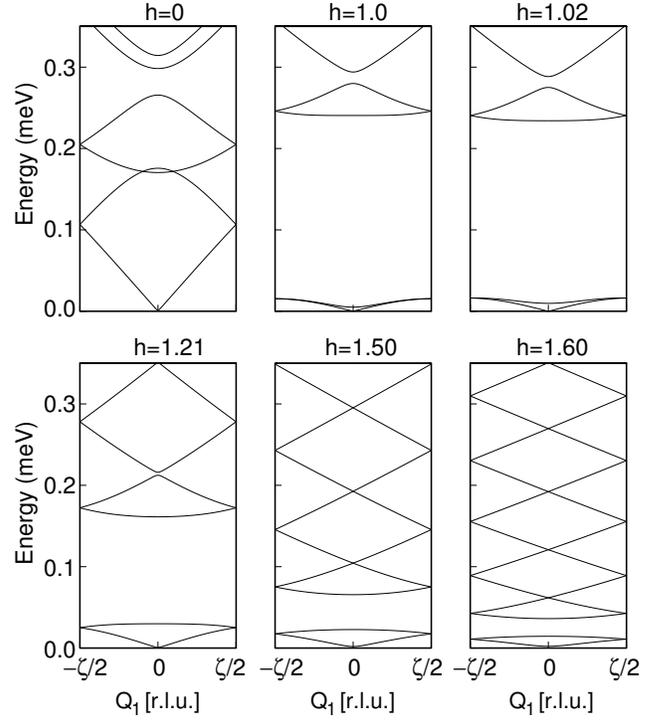} }
\caption{Magnon spectrum along the $x$ direction in the reduced zone scheme
for six illustrative magnetic fields. For $h>h_{c1}=1.01$ the results
show the magnon spectrum of the intermediate phase. 
The wave number $Q_{1}$ is measured in relative lattice units 
defined as $\zeta=\varepsilon/L=0.1774/L$.
Note that the lowest-lying band has a linear dispersion relation.}

\label{fig:12} 
\end{figure}

\begin{figure}
\resizebox{0.47\textwidth}{!}{\includegraphics{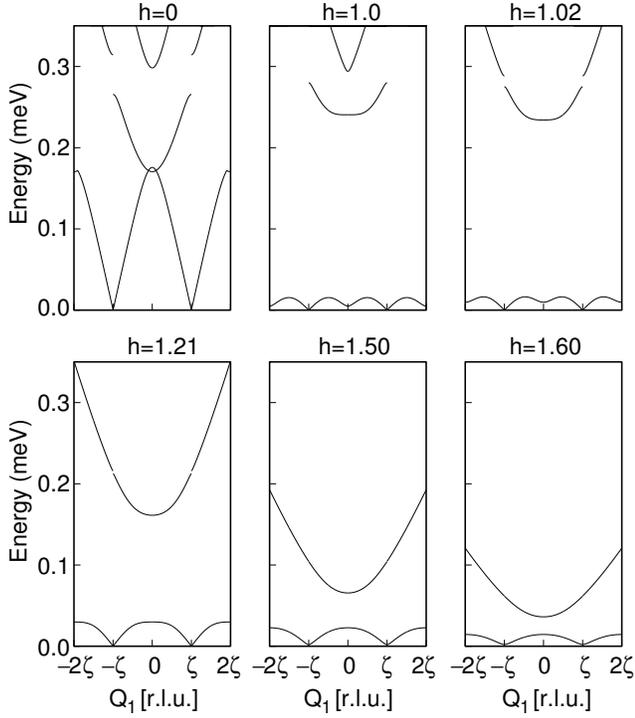} } 
\caption{Magnon spectrum along the $x$ direction in the extended zone scheme
for six illustrative magnetic fields. For $h>h_{c1}=1.01$ the results
show the magnon spectrum of the intermediate phase. Bands have been
assembled in a fashion that corresponds with conventions in publication
of experiments. }

\label{fig:13} 
\end{figure}

\begin{figure}[t]
\resizebox{0.47\textwidth}{!}{\includegraphics{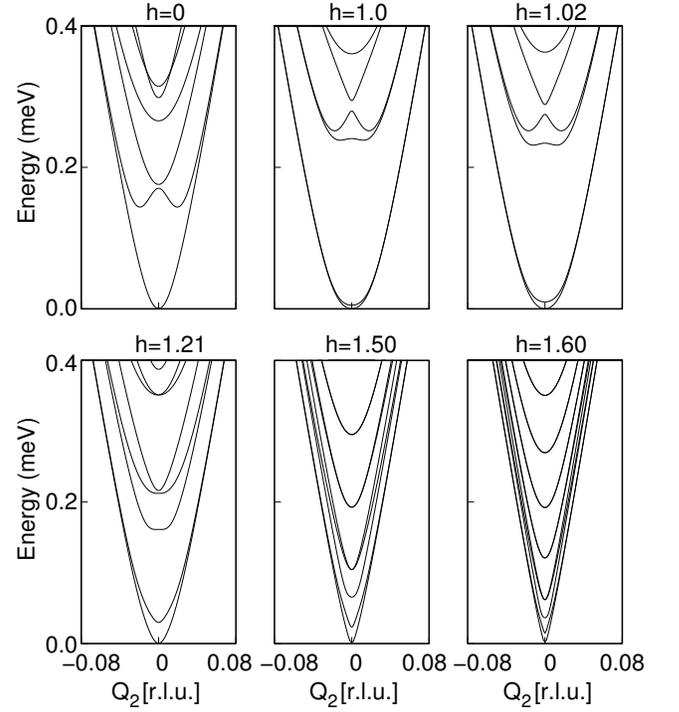} }
\caption{Magnon spectrum along the $y$ direction for six illustrative magnetic
fields. For $h>h_{c1}=1.01$ the results show the magnon spectrum
of the intermediate phase. 
The wave number $Q_{2}$ is measured in relative lattice units explained in the text.  
}

\label{fig:14} 
\end{figure}

We have solved the linear system (\ref{eq:2.15}) by a Bloch analysis
of the type given in Appendix A of Ref.~24 now extended to calculate
the low-energy magnon spectrum throughout the intermediate phase $h_{c1}<h<h_{c2}$.
The numerical procedure yields eigenfrequencies ${\omega}(q_1,q_2)$ 
as functions of Bloch momentum ${\bf q}=(q_1,q_2)$.
Since the potential terms on the right-hand side of Eq.~(\ref{eq:2.15})
are periodic along $x$ with period $L$, the component $q_1$
of the Bloch momentum can be restricted to the zone $[-\pi/L,\pi/L]$.
But $q_2$ is unrestricted because the background spin spiral 
is independent of $y$.

We present the results of the magnon calculations in Figs.~\ref{fig:12}
through \ref{fig:14}. 
The Bloch momentum in the figures is quoted in relative lattice units
${\bf Q}$[r.l.u.]$=(\varepsilon/2{\pi}){\bf q}=0.028{\bf q}$, following
conventions in publication of experiments.
Note that 
the value ${\bf Q}$[r.l.u.]$=1$ corresponds to 
Bloch wavelength of one lattice spacing  
of the square lattice formed by the Cu atoms within each layer.
The component $Q_1$ along the $x$ axis can now be restricted to
the zone $[-\zeta/2, \zeta/2]$, where $\zeta=\varepsilon/L=0.1774/L$.

We make the following comments: 
\begin{itemize}
\item All eigenvalues are positive. Therefore the intermediate state is
locally stable. This computation does not prove it is the ground state,
but in combination with extensive numerical explorations of two-dimensional
states that found no solutions of lower energy, it is a strong indication. 
\item We provide plots both in the reduced zone scheme and the extended
zone scheme. The reduced zone scheme is more compact, particularly
for $h_{c1}$ and below. However as the field increases towards $h_{c2}$
the reduced zone scheme acquires a large number of bands that are
resolved more clearly in the extended zone scheme. Experimentalists
are likely to find the display in the extended zone scheme more useful. 
\item Along $Q_{1}$ the low-energy spectrum is linear at the zone center.
Moving towards $h_{c2}$ it acquires two bands, an `acoustic' band
with linear dispersion and an upper optical band (higher bands exist
that have not been resolved by the computation). The linear portion
of the acoustic band is the Goldstone mode of these magnetic spin
states. In the limit that $h\rightarrow h_{c2}$ the bands depicted
here collapse onto the horizontal axis; the next excitation is at
an energy over 0.4 that lies above the top of the figure. 
\item Along $Q_{2}$ the low-energy spectrum is quadratic. As $h$ increases
towards $h_{c2}$ the quadratic regions become small and the spectrum
becomes nearly linear. Upon reaching $h_{c2},$ the dispersion becomes
completely linear. At this point it produces the Goldstone mode of
the spin-flop phase. 
\end{itemize}

\section{Local stability of the Spin-flop phase in canted magnetic fields}
\label{sec:local_stability}

Here we calculate the magnon spectrum of the Spin-flop phase in the presence
of canted magnetic field given by Eq.~(\ref{eq:2.3})
and thus examine an important issue concerning its stability.

We first note that the uniform Spin-flop state ${\bf n}=(-1,0,0)$,
or $\Phi=-\frac{\pi}{2}$, $\Theta=\frac{\pi}{2}$ using the spherical
parametrization (\ref{eq:2.3}), is an more or less obvious stationary
point that minimizes the energy functional $W=\int\, Vdx\, dy$, where
$V$ is the potential given in Eq.~(\ref{eq:4.2}). Actually, there
exist two different spin-flop configurations ${\bf n}=(\mp1,0,0)$
and both of them are the stationary points of the corresponding energy
functional. However, their energy densities given by $w=\frac{1}{2}\left(1\mp2h_{\perp}d_{z}\right)$
are different. Therefore, we will only consider the Spin-flop state
${\bf n}=(-1,0,0)$ with lower energy in our analysis. To examine
the stability, we first introduce new fields 
\begin{equation}
\Phi(x,y,t)=-\frac{\pi}{2}+f(x,y,t)\,,\,\,\,\Theta(x,y,t)=\frac{\pi}{2}+g(x,y,t)\,,\label{eq:4.4}
\end{equation}
%
where $f(x,y,t)$, $g(x,y,t)$ account for small fluctuations around
the Spin-flop state. Now the actual parametrization of the staggered
magnetization ${\bf n}$ given by Eq.~(\ref{eq:4.4}) is inserted in
the complete Lagrangian of Eq.~(\ref{eq:1.4}), which is applied for
a magnetic field ${\bf h}$ given by Eq.~(\ref{eq:4.1}) and expanded
to quadratic order in $f$, $g$. If we further perform the usual
Fourier transformation with frequency $\omega$ and wave vector ${\bf q}=(q_{1},q_{2})$,
the corresponding linearized equations of motion can be solved analytically
to yield the (squared) eigenfrequencies 
\begin{eqnarray}
{\omega}_{\pm}^{2}({\bf q}) & = & q_{1}^{2}+q_{2}^{2}+h_{\perp}d_{z}+\label{eq:4.5}
\end{eqnarray}
\[
+\frac{1}{2}\left(1+h_{z}^{2}+h_{\perp}^{2}\pm\sqrt{\left(1+h_{z}^{2}-h_{\perp}^{2}\right)^{2}+4h_{z}^{2}h_{\perp}^{2}+16q_{2}^{2}}\right)\,.
\]

The above calculated magnon spectrum is strongly anisotropic. To examine
the local stability of the Spin-flop state, we note that the stability
condition requires that $\omega_{+}^{2}\geq0$ and $\omega_{-}^{2}\geq0$
for each ${\bf q}$. It is also clear that $\omega_{+}^{2}\geq\omega_{-}^{2}$,
and $\omega_{-}^{2}$ is minimum for $q_{1}=0$. Therefore, we minimize
$\omega_{-}^{2}$ with respect to $q_{2}$ and then set $\omega_{-}^{2}=0$
to obtain 
\begin{equation}
h_{z}^{2}=3-h_{\perp}^{2}-2\sqrt{h_{\perp}^{2}+4h_{\perp}d_{z}}.\label{eq:4.6}
\end{equation}
The above obtained line of local stability of the Spin-flop state
is displayed by the dashed line in the phase diagram of Fig.~\ref{fig:4}.
Below the dashed line, the Spin-flop state is locally unstable and
cannot exist. 
Note that Eq.~(\ref{eq:4.6}) applied for the special case 
$h_{\perp}=0$ yields  $h_{z}={\sqrt{3}}$ ($H_{z}=2.9$ T), which is just the upper
critical field $h_{c2}$ obtained in Ref.~24.

\end{document}